\documentclass[gmd]{copernicus}

\usepackage{tabularx}

\usepackage{multirow}
\usepackage{supertabular}
\usepackage{algorithmic}
\usepackage{algorithm}
\usepackage{amsthm}
\usepackage{float}
\usepackage{subfig}
\usepackage[dvipsnames]{xcolor}
\usepackage{listings}
\usepackage{svg}
\definecolor{DarkBlue}{rgb}{0.00,0.00,0.55}
\definecolor{DarkRed}{rgb}{0.55,0.00,0.00}
\definecolor{DarkGreen}{rgb}{0.00,0.55,0.00}
\definecolor{Purple}{rgb}{0.5, 0.0, 0.5}
\definecolor{BurntOrange}{RGB}{204,85,0}
\definecolor{backcolour}{rgb}{0.95,0.95,0.92}

\lstdefinelanguage[firedrake]{python}[]{python}{%
   emph=[1]{UnitSquareMesh,PeriodicUnitIntervalMesh,SpatialCoordinate,VectorFunctionSpace,assemble,solve,FunctionSpace,Function,TestFunction,interpolate,Ensemble,NonlinearVariationalProblem,NonlinearVariationalSolver,VertexOnlyMesh,EnsembleFunctionSpace,EnsembleFunction,AutoregressiveCovariance,assign,sequential,WC4DVarReducedFunctional,record_stages,set_observation},
   emph=[2]{dx,inner,grad,dot,nabla_grad},
   emph=[3]{with,as,None,True},
   emph=[4]{continue_annotation,pause_annotation,Control,ReducedFunctional,TAOSolver,MinimizationProblem}
}

\lstdefinestyle{framed}{frame=tblr}

\usepackage{tikzit}

\tikzstyle{Mx}=[color=DarkBlue, line width=3pt, fill=white, -]
\tikzstyle{dash-Hx}=[draw={rgb,255: red,35; green,35; blue,51}, line width=1.5pt, dash dot, fill=white, <->]
\tikzstyle{dash-Mx}=[draw={rgb,255: red,35; green,35; blue,51}, line width=1.5pt, dashed, fill=white, <->]
\tikzstyle{dash-dt}=[draw={rgb,255: red,35; green,35; blue,51}, line width=1.0pt, densely dotted, fill=white, -]
\tikzstyle{rarrow}=[->, line width=3pt, fill=black]
\tikzstyle{tick}=[-, line width=3pt, fill=black]
\tikzstyle{marker}=[-, line width=1pt, fill=black]

\usepackage{multirow}

\begin{document}
\nolinenumbers

\title{fdvar: Automated parallel-in-time 4DVar data assimilation for any PDE using Firedrake}

\Author[1][joshua.hope-collins13@imperial.ac.uk]{Joshua}{Hope-Collins} 
\Author[2]{Jemima M.}{Tabeart}
\Author[1]{David A.}{Ham}
\affil[1]{Department of Mathematics, Imperial College London}
\affil[2]{Department of Mathematics and Computer Science, TU Eindhoven}

\runningtitle{4DVar with Firedrake}

\runningauthor{J. Hope-Collins, J. M. Tabeart and D. A. Ham}

\received{}
\pubdiscuss{} 
\revised{}
\accepted{}
\published{}

\firstpage{1}

\maketitle

\begin{abstract}
Data assimilation (DA) has made a significant contribution to the increase in forecast skill by leveraging real-world observations to improve the states used in the forecasting system. Four-dimensional variational assimilation (4DVar) is a commonly-used DA method at meteorological centres, where a nonlinear least squares problem is optimised in order to find the state that best fits previous forecast and observation information. This is achieved via the solution of a sequence of linearised system and adjoint equations, which represents a significant development bottleneck in applying 4DVar to new problems; manually differentiation requires high developer effort, while automatic numerical differentiation is computationally inefficient. 

We seek to overcome this obstacle via fdvar, a Python library which constructs the 4DVar system using symbolic automatic differentiation. 
This is achieved using: Firedrake, a finite element library which uses Unified Form Language to provide symbolic differentiation; and Pyadjoint, a library for symbolically recording code execution and automatically constructing the forward and adjoint models. To construct the 4DVar system, the user need only run the forward model and the observation operators once. Firedrake and Pyadjoint record this run and calculate the linearised forward and adjoint models, from which all components of 4DVar can be constructed. 

The resulting library offers full flexibility and high performance. Changes can readily be made to the data assimilation problem, including the underlying  model and observation network. Once these have been defined, no further code is required to construct the nonlinear objective function. Users have full control over solver options, including a variety of formulations and preconditioners, with the corresponding linear systems being automatically constructed by the library. In particular we have included both the primal and saddle point formulation of the weak constraint 4DVar problem, the latter of which offers inherently parallelisability across observation times. Finally, the implementation is both space and time parallel, enabling the real performance of the methods to be tested on HPC machines.

We present a step-by-step walkthrough of the code for an application to Burgers’ problem and illustrate the flexibility of the implementation for the nonlinear shallow water equations. 
Our new library fdvar closes the gap between small-scale toy problems and high-complexity operational-scale codes, allowing a smoother transition up the model hierarchy, and simplifies testing 4DVar methods on new applications. 
\end{abstract}

\introduction\label{sec:intro} 
Data assimilation methods combine observed data from a time-dependent physical system with information from a PDE model of that system, in order to obtain an improved estimate of the state at a given time.
Variational data assimilation methods represent a form of smoothing problem, where the full trajectory of the state is estimated over a specified time window.
While  there has been a large body of work applying variational data assimilation approaches to systems in geosciences, including atmosphere, ocean, hydrological and land surface systems (\cite{carrassiDataAssimilationGeosciences2018a}), they represent a general framework can be applied to many problems of scientific and societal interest.
Variational methods aim to solve a nonlinear least squares optimisation problem, which requires access to gradient and adjoint information of the underlying model.
This can be a major bottleneck in applying variational data assimilation methods to new applications and advancing methods and solvers.
This limits both the availability of test problems for traditional application areas, and the ease of trialling methodologies for novel applications. In comparison to ensemble data assimilation, there is a distinct lack of toolboxes for applying variational data assimilation to general problems \cite{Daužickaitė_Freitag_Gürol_Lawless_Ramage_Scott_Tabeart_2025}. To the best of our knowledge existing software that implements variational data assimilation methods automatically is restricted to 3DVar (\cite{ridlerDataAssimilationFramework2014,nergerAdvancingDataAssimilation2026}), small dimensional problems \cite{Raanes2024}, or requires explicit provision of adjoints and tangent linear models such as in OOPS (\cite{bonavita2017strategy}) and JEDI (\cite{tremolet2020joint}). Other testing suites  are often designed to interface with existing model code (e.g. \cite{attiaDATeSHighlyExtensible2019}).

Four-dimensioal variational data assimilation (4DVar) is one of the most widely used methods in practice, particularly in numerical weather prediction.
4DVar methods fall into two main categories.
``Strong constraint'' methods (SC4DVar) require the estimated state to be a perfect solution to the PDE model.
The focus of the current work is ``weak-constraint'' (WC4DVar) methods, where the estimated state is permitted to vary slightly from the PDE model if doing so brings it closer to the observations, acknowledging the fact that the PDE model is not a perfect representation of the real system.
Of particular importance for the current work, certain formulations of WC4DVar enable the optimisation problem to be solved parallel-in-time i.e. evaluating different parts of the timeseries simultaneously on different processors \citep{Fisher_Gürol_2017}.
Time-parallelism enables scaling beyond the limits of purely spatial parallelism, however it requires different solution methods to more traditional serial-in-time formulations.
Implementing time-parallelism also adds an extra layer of complexity to WC4DVar on top of that of implementing adjoint methods.

\subsection{Contributions}

In this work we present the inclusion of 4DVar methods within the Firedrake finite element library \citep{FiredrakeUserManual}.
Firedrake uses automatic differentiation to calculate the derivatives of PDE models, which means that once a user has implemented the forward model, the gradient and adjoint information is immediately available.
We leverage this information to automatically construct the 4DVar system based only on the user's forward model.
Enabling users to apply 4DVar to any PDE without having to write adjoint code has a number of benefits.
Firstly, existing DA users and developers can test their 4DVar methods, including sophisticated solvers and preconditioners, on a wider variety of PDEs, thus improving performance, robustness, and understanding of limitations and applicability.
In particular, this enables a smoother transition from simple toy models to more realistic but complex models when developing novel methods.
Secondly, researchers and application scientists from fields where DA is not yet standard can test out 4DVar methods on relevant PDE problems with minimal time investment, and without requiring prior expertise in adjoint methodology.

The major contributions of the current work are twofold.
\begin{enumerate}
    \item We have implemented, and designed a user friendly interface for immediate construction of the parallel-in-time 4DVar objective functional given minimal inputs from the user.
    \item We have automated the construction of the necessary matrices and appropriate preconditioners arising from the WC4DVar objective functional. 
\end{enumerate}
Other contributions include:
\begin{itemize}
    \item an interface for autoregressive covariance operators, 
    \item software components for time-parallel optimisation methods in general, 
    \item additions to the coupling with the PETSc and TAO optimisation library \cite{petsc-efficient,petsc-user-ref}. 
\end{itemize}

The paper is structured as follows.
In Section \ref{sec:methods} we introduce the variational data assimilation problem, and comment on general solution approaches and preconditioners.
In Section \ref{sec:firedrake} we describe some important features of the Firedrake library.
In Section \ref{sec:aims} we then describe at a high level what is necessary in order to apply 4DVar methods in Firedrake.
Sections  \ref{sec:covariance} and \ref{sec:obsoperators} describe two of these ingredients, respectively: the covariance operators and the observation operators.
We present a step-by-step walkthrough of the code for a motivating example applied to Burgers' equation in Section \ref{sec:software:demo}, and describe some specific implementation details in Section \ref{sec:software:impl}.
We illustrate the potential of the implementation in Section \ref{sec:examples} on {two PDEs motivated by geosciences; namely Burgers' equations and the nonlinear shallow water equations}. 
Finally in Section \ref{sec:conclusions} we present our conclusions.

\subsection{Notation}

In what follows lower case letters denote vectors, upper case letters denote matrices.
We use $N_t$ to denote the number of observation times in the observation window, which we will also refer to as the number of \textit{stages}, $N_{o_i}$ is the number of observations at time $t_i\in [0,T],i\in[0,N_t]$, and the total number of observations over the window is $N_o = \sum_{i=0}^{N_t}N_{o_i}$.
The size of the state at a single observation time is $N_x$.
Italic letters represent a single observation time, i.e. a spatial representation, such as $x_i, M_i$.
Bold letters represents a space-time object i.e. a collection of italic letters with one per observation e.g. $\mathbf{x} = [x_0,x_1,\dots,x_{N_t}]$.
Calligraphic font is used to denote nonlinear operators, e.g. $\mathcal{H}$.

Because we use finite elements to discretise in space, it is natural to consider each object to be a member of a function space, e.g. \(x\in V\).
Additionally, we distinguish the primal space \(V\), and its dual space \(V^{\star}\).
This distinction is essential for robust optimisation performance \citep{schwedes_mesh_2017}, however we do point out that an in depth understanding of the distinction is not necessary for understanding the current work.
Briefly, the dual space \(V^{\star}:V\to\mathbb{R}\) contains all linear functionals mapping an member \(x\in V\) to a real number in \(\mathbb{R}\).
For example, given a scalar function \(f(x)\), the derivative \(\nabla f\) is a member of \(V^{\star}\) because it maps a direction \(\delta x\in V\) to a change in the function \(\delta f\in\mathbb{R}\).
We point out a couple of consequences of using function space notations.
Firstly, given a linear operator \(L:V\to U\) from \(v\in V\) to \(u\in U\), where \(v\) has size \(N\) and \(u\) is size \(M\), there is an equivalent matrix of size \(\hat{L}\in\mathbb{R}^{M\times N}\).
Secondly, for any symmetric positive definite matrix, $A$, if we define a weighted norm of \(x\in V\)
\begin{equation}
	\|x\|_{A}^{2} = x^{T}Ax = x^{T}(Ax)
\end{equation}
then we can see that the action of \(Ax\) on \(x\) returns a real number i.e. \((Ax)\in V^{\star}\) and therefore \(A:V\to V^{\star}\).

\section{4D variational data assimilation}\label{sec:methods}
In this section we describe the 4D variational data assimilation problem, and introduce the notation and structure that underpin the software implementation.
More detailed derivations of this problem can be found in e.g. \cite{Daužickaitė_Freitag_Gürol_Lawless_Ramage_Scott_Tabeart_2025,lawDataAssimilationMathematical2015,aschDataAssimilation2016}.

Data assimilation methods seek to combine measurement information with prior information coming from an underlying dynamical system.
In variational assimilation, measurements are taken within a pre-specified time-window $[t_0,t_N]$, and the goal is to find an updated estimate of the state of the system that better matches both the measurements and prior information within the time-window.
All modelled fields are assumed to be Gaussian.

The solution to the 4DVar problem at observation time $t_i$ is denoted $x_i$, and its function space by \(V_{i}\).
\begin{equation}
	x_i \in V_{i},
	\quad
	\mathcal{M}_i: V_{i-1} \to V_{i}.
\end{equation}
For notational convenience we fix $V_{i}=V$ constant across all observation times from here on.
We note that what follows extends to $V_i$ varying at each observation time (e.g. with a temporally evolving mesh).
We assume that $x_i\in V$ satisfies the model $\mathcal{M}_i$
\begin{equation}
	x_{i} = \mathcal{M}_{i}\left(x_{i-1}\right) + \nu_{i},
	\quad
	\nu_{i}\sim\mathcal{N}\left(0,Q_{i}\right),
	\quad
	1\leq i \leq N.
\end{equation}
The model error covariance at observation time $t_i$ is denoted by $Q_i:V^{\star}\to V$, (i.e. there is an equivalent matrix \(Q_{i}\in\mathbb{R}^{N_{x}\times N_{x}}\)).
In practice each \(\mathcal{M}_{i}\) is often multiple timesteps of the numerical PDE model, so \(t_{i+1}-t_{i}\neq\Delta t\) in the usual sense of timestepping methods.
We denote the solution over the full time-window, also called the analysis, by $\mathbf{x} = [x_0, x_1,\dots,x_N] \in V^{N+1}=\mathbf{V}$.

The prior, $x_b \in V$, provides an estimate of  the state at $t_0$ and satisfies
\begin{equation}
	x_{b} = {x}_{0} + \varepsilon_{b},
	\quad
	\varepsilon_{b}\sim\mathcal{N}\left(0,B\right),
\end{equation}
where $B:V^{\star}\to V$ is called the background error covariance operator.

Observations \(y_i\) are made at time $t_i$ in observation space \(U_i\)
\begin{equation}
	y_i \in U_i,
	\quad
	\mathcal{H}_i: V \to U_i.
\end{equation}
The observation operator $\mathcal{H}_i$ maps from state space to observation space, and can vary between observation times.
The observations satisfy
\begin{equation}
	y_{i} = \mathcal{H}_i(x_i) + \mu_{i},
	\quad
	\mu_i\sim\mathcal{N}\left(0,R_{i}\right), \quad 0\le i \le N
\end{equation}
where $R_i: U_{i}^{\star}\to U_{i}$ is the observation error covariance operator at time $t_i$.
Similarly to \(\mathbf{V}\) we define the function space for the observations at all \(t_{i}\) as \(\mathbf{U}=\prod^{N}_{i=0}U_{i}\).

We refer to all operations and objects defined at a particular observation time \(i\), including \(\mathcal{M}_{i}\), \(\mathcal{H}_{i}\), \(Q_{i}\), \(R_{i}\), \(y_{i}\), and \(x_{i}\) as belonging to the same \emph{observation stage}.

Variational data assimilation methods seek a solution \(\mathbf{x}\) which satisfies \(\mathcal{M}\) (possibly approximately) and minimises the deviation from the prior \(x_{b}\) and the mismatch between the real world observations \(y_i\) and the numerical observations \(\mathcal{H}_i(x_i)\).
We now introduce the numerical problem formulation.

\subsection{Weak constraint 4DVar}\label{sec:methods:wc4DVar}
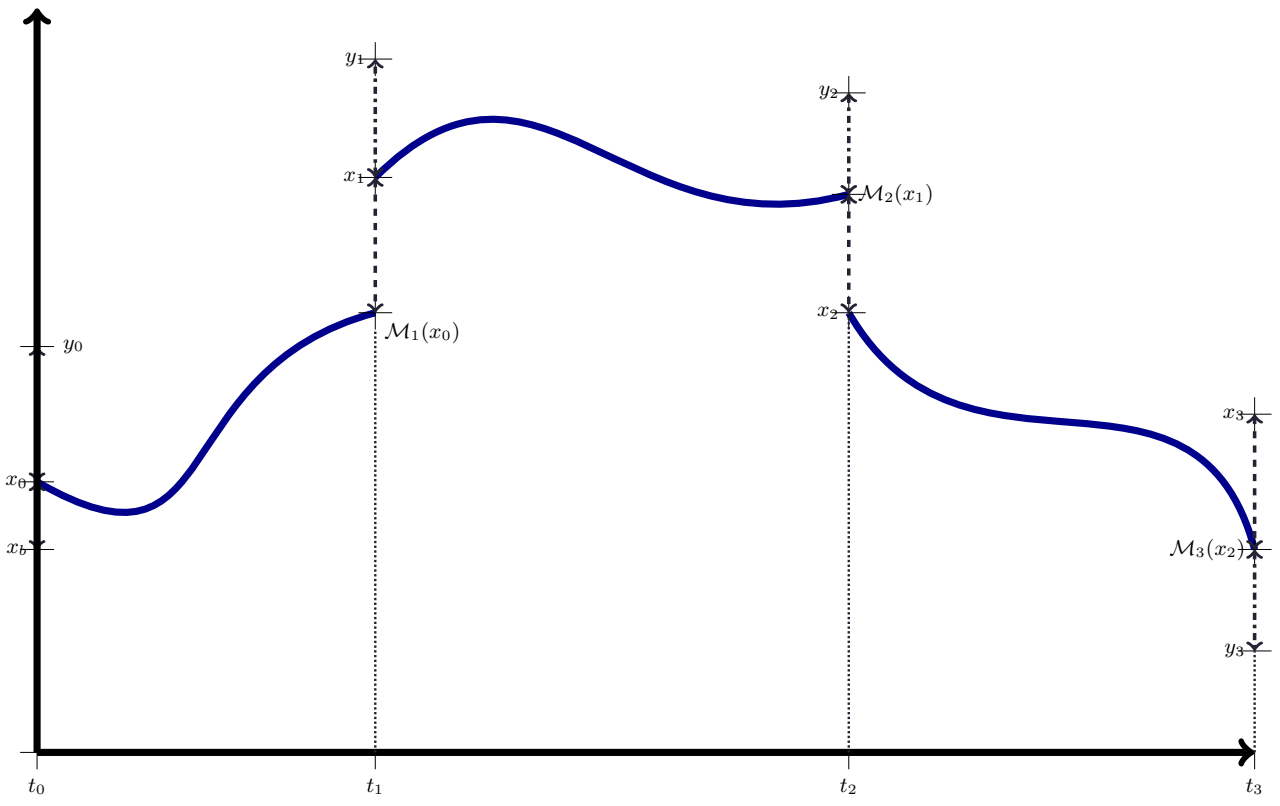
\begin{figure}
	\hspace{-3em}\resizebox{\linewidth}{!}{
			\noindent\begin{tikzpicture}[baseline]
				\begin{tikzpicture}
	\begin{pgfonlayer}{nodelayer}
		\node [style=none] (1) at (0, 0) {};
		\node [style=none] (2) at (18, 0) {};
		\node [style=none] (3) at (0, 11) {};
		\node [style=none, label={left:\(x_{b}\)}] (4) at (0, 3) {};
		\node [style=none, label={left:\(x_{0}\)}] (5) at (0, 4) {};
		\node [style=none] (6) at (0, 6) {};
		\node [style=none, label={below right:\(\mathcal{M}_{1}(x_{0})\)}] (10) at (5, 6.5) {};
		\node [style=none, label={left:\(y_{1}\)}] (11) at (5, 10.25) {};
		\node [style=none, label={left:\(x_{1}\)}] (12) at (5, 8.5) {};
		\node [style=none, label={right:\(\mathcal{M}_{2}(x_{1})\)}] (13) at (12, 8.25) {};
		\node [style=none, label={left:\(x_{2}\)}] (14) at (12, 6.5) {};
		\node [style=none, label={left:\(y_{2}\)}] (15) at (12, 9.75) {};
		\node [style=none, label={left:\(\mathcal{M}_{3}(x_{2})\)}] (16) at (18, 3) {};
		\node [style=none, label={left:\(x_{3}\)}] (17) at (18, 5) {};
		\node [style=none, label={left:\(y_{3}\)}] (18) at (18, 1.5) {};
		\node [style=none] (19) at (5, 10.5) {};
		\node [style=none] (20) at (5, 10) {};
		\node [style=none] (21) at (5.25, 10.25) {};
		\node [style=none] (22) at (4.75, 10.25) {};
		\node [style=none] (23) at (5, 8.75) {};
		\node [style=none] (24) at (5.25, 8.5) {};
		\node [style=none] (25) at (5, 8.25) {};
		\node [style=none] (26) at (4.75, 8.5) {};
		\node [style=none] (27) at (5, 6.75) {};
		\node [style=none] (28) at (5.25, 6.5) {};
		\node [style=none] (29) at (5, 6.25) {};
		\node [style=none] (30) at (4.75, 6.5) {};
		\node [style=none] (31) at (12, 10) {};
		\node [style=none] (32) at (12.25, 9.75) {};
		\node [style=none] (33) at (12, 9.5) {};
		\node [style=none] (34) at (11.75, 9.75) {};
		\node [style=none] (35) at (12, 8.5) {};
		\node [style=none] (36) at (12.25, 8.25) {};
		\node [style=none] (37) at (12, 8) {};
		\node [style=none] (38) at (11.75, 8.25) {};
		\node [style=none] (39) at (12, 6.75) {};
		\node [style=none] (40) at (11.75, 6.5) {};
		\node [style=none] (41) at (12.25, 6.5) {};
		\node [style=none] (42) at (12, 6.25) {};
		\node [style=none] (44) at (17.75, 1.5) {};
		\node [style=none] (45) at (18, 1.25) {};
		\node [style=none] (46) at (18.25, 1.5) {};
		\node [style=none] (47) at (18, 2.75) {};
		\node [style=none] (48) at (18.25, 3) {};
		\node [style=none] (49) at (17.75, 3) {};
		\node [style=none] (50) at (18, 3.25) {};
		\node [style=none] (51) at (18, 5.25) {};
		\node [style=none] (52) at (17.75, 5) {};
		\node [style=none] (53) at (18.25, 5) {};
		\node [style=none] (54) at (18, 4.75) {};
		\node [style=none] (55) at (0.25, 4) {};
		\node [style=none] (56) at (0.25, 3) {};
		\node [style=none, label={right:\(y_{0}\)}] (57) at (0.25, 6) {};
		\node [style=none] (58) at (-0.25, 3) {};
		\node [style=none] (59) at (-0.25, 4) {};
		\node [style=none] (60) at (-0.25, 6) {};
		\node [style=none, label={below:\(t_{0}\)}] (61) at (0, -0.25) {};
		\node [style=none, label={below:\(t_{1}\)}] (62) at (5, -0.25) {};
		\node [style=none, label={below:\(t_{2}\)}] (63) at (12, -0.25) {};
		\node [style=none, label={below:\(t_{3}\)}] (64) at (18, -0.25) {};
		\node [style=none] (65) at (5, 0) {};
		\node [style=none] (66) at (12, 0) {};
		\node [style=none] (67) at (-0.25, 0) {};
	\end{pgfonlayer}
	\begin{pgfonlayer}{edgelayer}
		\draw [style=rarrow] (1.center) to (2.center);
		\draw [style=Mx, in=-165, out=-30, looseness=1.50] (5.center) to (10.center);
		\draw [style=Mx, in=-165, out=45, looseness=1.25] (12.center) to (13.center);
		\draw [style=Mx, in=105, out=-60, looseness=1.25] (14.center) to (16.center);
		\draw (51.center) to (54.center);
		\draw (52.center) to (53.center);
		\draw (50.center) to (47.center);
		\draw (49.center) to (48.center);
		\draw (44.center) to (46.center);
		\draw (31.center) to (33.center);
		\draw (34.center) to (32.center);
		\draw (35.center) to (37.center);
		\draw (38.center) to (36.center);
		\draw (39.center) to (42.center);
		\draw (40.center) to (41.center);
		\draw (19.center) to (20.center);
		\draw (22.center) to (21.center);
		\draw (23.center) to (25.center);
		\draw (26.center) to (24.center);
		\draw (27.center) to (29.center);
		\draw (30.center) to (28.center);
		\draw (60.center) to (57.center);
		\draw (59.center) to (55.center);
		\draw (58.center) to (56.center);
		\draw [style=dash-Mx] (12.center) to (10.center);
		\draw [style=dash-Hx] (12.center) to (11.center);
		\draw [style=dash-Hx] (15.center) to (13.center);
		\draw [style=dash-Mx, in=90, out=-90] (13.center) to (14.center);
		\draw [style=dash-Hx] (16.center) to (18.center);
		\draw [style=dash-Mx] (17.center) to (16.center);
		\draw [style=dash-Hx] (5.center) to (6.center);
		\draw [style=dash-Mx] (5.center) to (4.center);
		\draw (2.center) to (64.center);
		\draw (66.center) to (63.center);
		\draw (65.center) to (62.center);
		\draw (1.center) to (61.center);
		\draw (1.center) to (67.center);
		\draw [style=rarrow] (1.center) to (3.center);
		\draw [style=dash-dt] (65.center) to (29.center);
		\draw [style=dash-dt] (66.center) to (14.center);
		\draw [style=dash-dt] (2.center) to (18.center);
	\end{pgfonlayer}
\end{tikzpicture}
			\end{tikzpicture}
		}
		\caption{Visualisation of weak constraint 4DVar solution}
		\label{methods:fig:wc4DVar}
\end{figure}

Weak constraint 4DVar (WC4DVar) methods account for the fact that the model is not perfect, i.e. \(Q_{i}\neq 0\).
Hence the solution $\mathbf{x}$ to WC4DVar only satisfies the PDE, $\mathcal{M}_i$, approximately.
This means that \(x_{i}\) is no longer solely a function of \(x_{0}\), as can be seen in the jumps in the state shown in Figure \ref{methods:fig:wc4DVar}.
The control for the optimisation is thus given by \(\mathbf{x}\in \mathbf{V}\).

The Gaussian assumption, and application of Bayes' Theorem, allows us to write the weak constraint 4DVar problem as the following objective function
\begin{equation}\label{methods:eq:wc-jhat}
	\begin{split}
		\mathcal{J}_{w}(\mathbf{x}) =\|x_{0} - x_{b}\|_{B^{-1}}^{2}
	& + \sum^{N}_{i=0} \|\mathcal{H}_{i}\left(x_{i}\right) - y_{i}\|_{R_{i}^{-1}}^{2} \\
		& + \sum^{N}_{i=1} \|x_{i} - \mathcal{M}_{i}\left(x_{i-1}\right)\|_{Q_{i}^{-1}}^{2}.
	\end{split}
\end{equation}
We note the three terms corresponding to fit to prior, fit to observations, and fit given the imperfect model assumption.

{Let $\mathbf{\mathcal{L}}:\mathbf{V}\rightarrow \mathbf{V}$ and $\mathbf{\mathcal{H}}:\mathbf{V}\rightarrow \mathbf{U}$ such that 
\begin{align}
	\mathbf{\mathcal{L}}(\mathbf{x}) & =
	\begin{pmatrix}
		x_0 - x_b \\
		x_1 - \mathcal{M}_1(x_0)       \\
		\vdots     \\
		x_N  - \mathcal{M}_N(x_{N-1}) \\
	\end{pmatrix} = -\mathbf{b},\\ \mathbf{\mathcal{H}}(\mathbf{x}) & =
	\begin{pmatrix}
		\mathcal{H}_0(x_0) - y_0 \\
		\mathcal{H}_1(x_1) - y_1       \\
		\vdots     \\
		\mathcal{H}_N(x_N) - y_N \\
	\end{pmatrix} = -\mathbf{d}.
    \end{align}
    This allows us to write the nonlinear objective function as 
    
    \begin{equation}\label{methods:eq:wc-LH}
        \mathcal{J}_{w}(\mathbf{x}) =\|\mathbf{\mathcal{L}}(\mathbf{x})\|_{\mathbf{D}^{-1}}^{2} +\|\mathbf{\mathcal{H}}(\mathbf{x})\|_{\mathbf{R}^{-1}}^{2}, 
    \end{equation}
    
    where 
    \begin{align}
	\mathbf{D} & =
	\begin{pmatrix}
		B &       &     &     \\
		& Q_{1} &     &       \\
		&       & \ddots &    \\
		&       &     & Q_{N} \\
	\end{pmatrix}
   \in \mathbf{V}^{\star} \to \mathbf{V} \label{methods:eq:D-block}\\
	\mathbf{R} & =
	\begin{pmatrix}
		R_{0} &       &     &       \\
		& R_{1} &     &       \\
		&       & \ddots &       \\
		&       &     & R_{N} \\
	\end{pmatrix}
   \in \mathbf{U}^{\star} \to \mathbf{U}. \label{methods:eq:R-block}
   \end{align}
   denote the concatenated covariance matrices. 

   In practice (truncated) Gauss-Newton  is often used to solve the nonlinear problem in Eq.~(\ref{methods:eq:wc-jhat}) \citep{nocedalNumericalOptimization2006a,grattonApproximateGaussNewton2007}.
At the $l$th iteration of Gauss-Newton, we aim to solve for \(\mathbf{\delta x}\)
\begin{align}
		\mathcal{J}_{w}^{(l)}(\mathbf{x}^{(l)} + \delta \mathbf{x}^{(l)})
		& =\|\mathbf{\mathcal{L}}(\mathbf{x}^{(l)} + \delta \mathbf{x}^{(l)})\|_{\mathbf{D}^{-1}}^{2} \\&+\|\mathbf{\mathcal{H}}(\mathbf{x}^{(l)} + \delta \mathbf{x}^{(l)})\|_{\mathbf{R}^{-1}}^{2}\nonumber.
\end{align}

We define the linearisations by
\begin{align}
\mathbf{\mathcal{L}}(x^{(l)} + \delta x^{(l)}) = \mathbf{\mathcal{L}}(x^{(l)}) + \mathbf{L}\delta x^{(l)} \\
\mathbf{\mathcal{H}}(x^{(l)} + \delta x^{(l)}) = \mathbf{\mathcal{H}}(x^{(l)}) + \mathbf{H}\delta x^{(l)} \\
	\mathcal{H}_{i}(x_{i}^{(l)} + \delta x_{i}^{(l)}) = \mathcal{H}_{i}(x_{i}^{(l)}) + H_{i}\delta x_{i}^{(l)} \\
	\mathcal{M}_{i}(x_{i-1}^{(l)} + \delta x_{i-1}^{(l)}) = \mathcal{M}_{i}(x_{i-1}^{(l)}) + M_{i}\delta x_{i-1}^{(l)}.
\end{align}
We note that we can write $\mathbf{L}$ and $\mathbf{H}$ explicitly as
\begin{align}
\mathbf{H} & = 
	\begin{pmatrix}
		H_{0} &       &     &       \\
		& H_{1} &     &       \\
		&       & \ddots &       \\
		&       &     & H_{N} \\
	\end{pmatrix}
   \in \mathbf{V} \to \mathbf{U} \label{methods:eq:H-block}\\
	\mathbf{L} & =
	\begin{pmatrix}
		I      &        &        &   \\
		-M_{1} & I      &        &   \\
		& \ddots    & \ddots    &   \\
		&        & -M_{N} & I \\
	\end{pmatrix}
   \in \mathbf{V} \to \mathbf{V}. \label{methods:eq:L-block}
\end{align}

This results in the linear system for the Gauss-Newton update \(\delta\mathbf{x}^{(l)}\):
\begin{align}\label{methods:eq:LLSweak}
	 J_{w}^{(l)}(\delta \mathbf{x}^{(l)})
	& = \|\mathbf{L}\mathbf{\delta x}  - \mathbf{b} \|_{ \mathbf{D}^{-1}}^2 + \|\mathbf{H}\mathbf{\delta x}  - \mathbf{d} \|_{\mathbf{R}^{-1}}^2.
\end{align}
In what follows we will omit the $(l)$ index to improve readability.
}

We note that Eq.~(\ref{methods:eq:LLSweak}) has inherently parallel-in-time structure, as seen from the fact that the action of its component matrices, Eq.~(\ref{methods:eq:H-block}), Eq.~(\ref{methods:eq:L-block}) can be applied at different observation times in parallel.
Matrices of the form of \(\mathbf{L}\) are often referred to as ``all-at-once'' systems in the parallel-in-time literature.
We note that many of these `matrices' are only available as the action of a matrix on a vector, in particular the linearised operators $M_i: V\to V$ and $H_i: V\to U_{i}$.
Therefore, any solvers (including preconditioners) discussed in this paper will not require explicit access to a matrix, but rather function with only matrix-vector products.
The computational cost of solving the weak constraint problem is typically dominated by the cost of operations with $\mathcal{M}_i$ and its linearisation $M_i$.

\subsection{Strong constraint 4DVar}\label{sec:methods:sc4DVar}
Strong constraint 4DVar (SC4DVar) assumes a perfect model, i.e. $x_i = \mathcal{M}_i(x_{i-1})$, or $Q_i = 0 \forall i$.
This leads to a simplification of the objective function and underlying problem compared to weak constraint 4DVar.
Namely, \(x_{i}\) is solely a function of \(x_{0}\), meaning that an equivalent of Figure \ref{methods:fig:wc4DVar} would have a continuous line over the observation time window.
This implies that the  control for the optimisation is \(x_{0}\in V\); we solve a significantly smaller dimensional problem than for the weak constraint variant.

The corresponding nonlinear least squares objective function is given by
\begin{align}\label{methods:eq:NLSstrong}
   \mathcal{J}_{s}(x_0) =
       \|x_{0} - x_{b}\|_{B^{-1}}^{2}
       + \sum^{N}_{i=0} \|\mathcal{H}_{i}\left(\mathcal{F}_{i}\left(x_{0}\right)\right) - y_{i}\|_{R^{-1}}^{2}.
\end{align}
Here $\mathcal{F}_i(x_0) = \mathcal{M}_i(\dots(\mathcal{M}_1(x_0)))$ propagates the state from $t_0$ to $t_i$. 
We note that Eq.~(\ref{methods:eq:NLSstrong}) contains only two terms corresponding to the prior fit, and fit to observations.

The Gauss-Newton iteration for \(\delta x_0^{(l)} = x_0^{(l+1)}-x_0^{(l)}\) is given by
\begin{align}
   J_{s}(x_{0}^{(l)} + \delta x_{0}^{(l)}) =
      & \|x_{0}^{(l)} + \delta x_{0}^{(l)} - x_{b}\|_{B^{-1}}^{2} \\
      & + \sum^{N}_{i=0} \|\mathcal{H}_{i}\left(\mathcal{F}_{i}\left(x_{0}^{(l)} + \delta x_{0}^{(l)}\right)\right) - y_{i}\|_{R^{-1}}^{2}.
\end{align}

Using the same notation as introduced in Section \ref{sec:methods:wc4DVar}, this linear least squares objective function can hence be written compactly as
\begin{equation}\label{methods:eq:LLSstrong}
	J_{s}(\delta x_{0}) = \|\delta x_0-b_0\|_{B^{-1}}^{2} + \|\mathbf{H}\mathbf{F}\delta x_0-\mathbf{d}\|_{\mathbf{R}^{-1}}^2
\end{equation}
where the Jacobian matrix $\mathbf{F}: V\to\mathbf{V}$ represents a concatenation of the linearised $F_i$ operators (the model $\mathcal{F}_i$ linearized about the current iterate).
We note that 
\begin{equation}\label{methods:eq:defineF}
    \mathbf{F}(x_0) = \mathbf{L}^{-1}\mathbf{E}_0x_0, 
\end{equation}
where $\mathbf{E}_0 =[I,\mathbf{0},\dots,\mathbf{0}]^T.$

\subsection{Solving the linear least squares problem}\label{sec:methods:lsq}
In this section we briefly discuss how to solve the linear least squares problems Eq.~(\ref{methods:eq:LLSweak}) and Eq.~(\ref{methods:eq:LLSstrong}).
Further detail can be found in \cite{Daužickaitė_Freitag_Gürol_Lawless_Ramage_Scott_Tabeart_2025}.
We aim to provide an overview of the underpinning structures of the problem that will be exploited in the computational implementation, as well as highlighting some challenges and opportunities for practical settings.

The linear least squares problem corresponding to the weak constraint formulation is  usually solved in one of two ways.
Most common is via the normal equations,
\begin{equation}\label{methods:eq:NEweak}
	\mathbf{S}\delta \mathbf{x} = \mathbf{L}^T\mathbf{D}^{-1}\mathbf{b}+
	\mathbf{H}^T\mathbf{R}^{-1}\mathbf{d},
\end{equation}
where
\begin{equation}\label{methods:eq:Hessianweak}
	\mathbf{S} = \mathbf{L}^T\mathbf{D}^{-1}\mathbf{L} + \mathbf{H}^T\mathbf{R}^{-1}\mathbf{H},
\end{equation}
is the Hessian of Eq.~(\ref{methods:eq:LLSweak}).
We refer to Eq.~(\ref{methods:eq:NEweak}) as the primal formulation.
This has the advantage that $\mathbf{S}$ is symmetric positive definite, meaning that Eq.~(\ref{methods:eq:NEweak}) can be solved using the conjugate gradient method (CG).
We note that $J_w$ and the action of $\mathbf{S}$ can be evaluated parallel-in-time by integrating over each observation time, or through each stage, forward and backward independently.
This can offer significant speed ups over serial evaluations.

A change of variable, $\mathbf{v} = \mathbf{L}\delta \mathbf{x}$, allows us to solve the so-called \textit{forcing formulation} via 
\begin{equation}\label{methods:eq:ForcingForm}
   ( \mathbf{D}^{-1} + \mathbf{L}^{-T}\mathbf{H}^T\mathbf{R}^{-1}\mathbf{H}\mathbf{L}^{-1})\mathbf{v} = \mathbf{D}^{-1}\mathbf{b} + \mathbf{L}^{-T}\mathbf{H}\mathbf{R}^{-1}\mathbf{d}.
\end{equation}
This is equivalent to solving Eq.~(\ref{methods:eq:NEweak}) with the preconditioners $\mathbf{L}^{-T}$ applied on the left, and $\mathbf{L}^{-1}$ applied on the right.

In order to accelerate the convergence of CG, preconditioners based on the first term of the Hessian \(\mathbf{\hat{S}}=\mathbf{L}^{T}\mathbf{D}^{-1}\mathbf{L}\) are usually applied.
If $\mathbf{L}^{-1}$ and $\mathbf{D}^{-1/2}$ are available, we can apply the this preconditioner symmetrically to obtain
\begin{equation}\label{methods:eq:symmetricL1wc}
	\mathbf{D}^{1/2}\mathbf{L}^{-T}\mathbf{S}\mathbf{L}^{-1}\mathbf{D}^{1/2}
	= \mathbf{I} + \mathbf{D}^{1/2}\mathbf{L}^{-T}\mathbf{H}^T\mathbf{R}^{-1}\mathbf{H}\mathbf{L}^{-1}\mathbf{D}^{1/2}.
\end{equation}
In this case, as the observation term is low-rank and semi-definite, there are $N_{t}N_{x} - N_{o}$ unit eigenvalues and the preconditioned system has provably good spectral results
\begin{equation}
	\sigma\left(\mathbf{D}^{1/2}\mathbf{L}^{-T}\mathbf{S}\mathbf{L}^{-1}\mathbf{D}^{1/2}\right) \geq 1.
\end{equation}
Both of these theoretical properties are linked to accelerated convergence of conjugate gradient.

However, the direct inversion $\mathbf{L}^{-1}$, which also appears in the forcing formulation \eqref{methods:eq:ForcingForm}, is equivalent to integrating over the entire timeseries. This requires serial evaluations of products involving $M_i$ operators and therefore cannot be parallelised over observation times.
It is also possible to use more computationally affordable approximations $\mathbf{\tilde{L}}\approx\mathbf{L}$ and $\mathbf{\tilde{D}}\approx\mathbf{D}$ to construct the preconditioner \(\mathbf{\tilde{S}}=\mathbf{\tilde{L}}^{T}\mathbf{\tilde{D}}^{-1}\mathbf{\tilde{L}}\approx\mathbf{\hat{S}}\).
In this case the beneficial eigenvalue properties no longer hold. The preconditioned system can still have problematic large eigenvalues which lead to delayed convergence, although improved performance may still be observed in practice.
Any approximation \(\mathbf{\tilde{L}}^{-1}\) should be much cheaper to apply that \(\mathbf{L}^{-1}\), either by using some approximation \(\tilde{M}_{i}\) which is simpler than \(M_{i}\), or by changing the structure such that it can be applied parallel-in-time \citep{gander_50_2015}.
If the preconditioner \(\mathbf{\tilde{S}}\) is applied symmetrically, as in Eq.~(\ref{methods:eq:symmetricL1wc}), then second-level preconditioners can also be applied to further improve performance (\cite{Daužickaitė_Lawless_Scott_Van_Leeuwen_2021}).

However, the convergence of the primal formulation is heavily dependent on a very good approximation \(\mathbf{\tilde{L}}\) \citep{Gratton_Gürol_Simon_Toint_2018a}, which motivates the second problem formulation as an augmented system or saddle point problem.
In this case we reformulate Eq.~(\ref{methods:eq:Hessianweak}) using the KKT conditions, leading to the following saddle point system
\begin{equation}\label{methods:eq:saddlepoint}
	\underbrace{\begin{pmatrix}
		\mathbf{D}     & \mathbf{0}     & \mathbf{L} \\
		\mathbf{0}     & \mathbf{R}     & \mathbf{H} \\
		\mathbf{L}^{T} & \mathbf{H}^{T} & \mathbf{0}
	\end{pmatrix}}_{\mathbf{A}}
	\begin{pmatrix}
		\mathbf{\eta} \\ \mathbf{\lambda} \\ \delta\mathbf{x}
	\end{pmatrix}
	=
	\begin{pmatrix}
		\mathbf{b} \\ \mathbf{d} \\ \mathbf{0}
	\end{pmatrix}
\end{equation}
where \(\nu\in\mathbf{V}^{\star}\) and \(\lambda\in\mathbf{U}^{\star}\) are Lagrange multipliers.

This saddle-point formulation can be applied parallel-in-time, and offers a number of advantages over the symmetric positive definite formulation.
There is a wealth of literature for preconditioning saddle point systems that can be leveraged to improve 4DVar performance.
The block triangular, and block diagonal preconditioners based on a Schur complement factorisation
\begin{align}
	\mathbf{P}_{U} & =
	\begin{pmatrix} 
		\mathbf{D} & \mathbf{0} & \mathbf{L} \\
		\mathbf{0} & \mathbf{R} & \mathbf{H} \\
		\mathbf{0} & \mathbf{0} & \mathbf{{S}}
	\end{pmatrix} \label{methods:eq:PU}\\
	\mathbf{P}_{L} & =
	\begin{pmatrix}
		\mathbf{D}     & \mathbf{0}     & \mathbf{0} \\
		\mathbf{0}     & \mathbf{R}     & \mathbf{0} \\
		\mathbf{L}^{T} & \mathbf{H}^{T} & \mathbf{{S}}
	\end{pmatrix} \label{methods:eq:PL} \\
	\mathbf{P}_{D} & =
	\begin{pmatrix}
		\mathbf{D} & \mathbf{0} & \mathbf{0} \\
		\mathbf{0} & \mathbf{R} & \mathbf{0} \\
		\mathbf{0} & \mathbf{0} & \mathbf{{S}}
	\end{pmatrix} \label{methods:eq:PD}
\end{align}
have been well-studied in the numerical linear algebra literature, see e.g. the review papers \cite{benziNumericalSolutionSaddle2005,wathenPreconditioning2015}.
We note that $\mathbf{S}$, the Schur complement of $\mathbf{A}$, is exactly the Hessian of the linear least squares problem.

Motivated by the desire to avoid products with $\mathbf{D}^{-1}$, which may not be available as an operator, \cite{Fisher_Gürol_2017} proposed the inexact constraint preconditioner:
\begin{align}
	\mathbf{P}_{C} & =
	\begin{pmatrix}
		\mathbf{D}     & \mathbf{0} & \mathbf{L} \\
		\mathbf{0}     & \mathbf{R} & \mathbf{0} \\
		\mathbf{L}^{T} & \mathbf{0} & \mathbf{0}
	\end{pmatrix}
\end{align}

Within each of these larger preconditioning structures, further approximations are typically made, particularly to $\mathbf{S}$.
For example, by a suitable choice of approximation to $\mathbf{L}$, the parallel-in-time structure of $\mathbf{A}$ can also be inherited by the preconditioners $\mathbf{P}_U,\mathbf{P}_L,\mathbf{P}_D$ and $\mathbf{P}_C$.
Therefore there is more scope to accelerate performance of the saddle point formulation while maintaining parallelisability than the symmetric positive definite formulation.
In what follows we use  \(\mathbf{\tilde{L}}\) (respectively and \(\mathbf{\tilde{D}}\)) to denote approximations to the matrices  \(\mathbf{{L}}\) and \(\mathbf{{D}}\) within preconditioners.

The linear least squares problem corresponding to the strong constraint formulation is usually solved via the normal equations,

\begin{equation}\label{methods:eq:primalstrong}
	S_{c}\delta x_0 = \mathbf{F}^{T}\mathbf{H}^T\mathbf{R}^{-1}\mathbf{d} + B^{-1}b_{0}
\end{equation}
where
\begin{equation}\label{methods:eq:Hessianstrong}
	S_{c} = B^{-1} + \mathbf{F}^T\mathbf{H}^T\mathbf{R}^{-1}\mathbf{H}\mathbf{F}
\end{equation}
is the Hessian of Eq.~(\ref{methods:eq:LLSstrong}), and $\mathbf{F}$ is defined in Eq.~(\ref{methods:eq:defineF}).

Similar to the weak constraint problem, preconditioning Eq.~(\ref{methods:eq:Hessianstrong}) with its first term leads to beneficial spectral properties;
\begin{equation}\label{methods:eq:splitprecond}
	B^{1/2}S_{c}B^{1/2} = I + B^{1/2}\mathbf{F}^{T}\mathbf{H}^T\mathbf{R}^{-1}\mathbf{H}\mathbf{F}B^{1/2} 
\end{equation}
is a low-rank update to the identity because \(\mathbf{H}:\mathbf{V}\to\mathbf{U}\) is low rank, and
\begin{equation}
	\sigma\left(B^{1/2}S_{c}B^{1/2}\right) \geq 1.
\end{equation}
We note that applying the \textit{split} preconditioning to the strong (respectively weak) constraint Hessian requires access to (a good approximation to) \(B^{1/2}\) (respectively $\mathbf{D}^{1/2}$).
This is  typically taken into account when designing/constructing $B$ (see Section \ref{sec:covariance}), but could be limiting for some applications.
If \(B^{1/2}\) (respectively \(\mathbf{D}^{1/2}\)) is not available then left preconditioning CG with \(B\) (respectively \(\mathbf{D}\)) is equivalent to split preconditioning, although it precludes the use of second-level preconditioning.

In contrast to the weak constraint problem, evaluating \(J_{s}\) requires integrating through entire timeseries, and is thus not inherently parallelisable-in-time.
Similarly, evaluating the action of Hessian requires integrating the TLM/adjoint forward/backward through entire timeseries.

 For traditional data assimilation applications in numerical weather prediction, the time allocated to ``solve'' the objective function is extremely limited.
This can often result in the application of truncated iterative methods with very limited budgets.
Other techniques to improve efficiency include evaluating linear solves on coarser meshes and dropping terms from the adjoint and TLM models, in addition to  the use of preconditioners.

\section{Firedrake finite element library}\label{sec:firedrake}
In this section we give a brief description of Firedrake (\cite{FiredrakeUserManual}).
Firedrake enables users to specify finite element problems in Python at a high level of abstraction using the Unified Form Language DSL \citep{alnaes_unified_2014}, and uses automatic code generation to compile high perfomance C kernels for carrying out the numerical calculations.
It integrates tightly with the PETSc library \citep{petsc-efficient,petsc-user-ref} via petsc4py, \citep{dalcinpazklercosimo2011} giving users access to a wide range of composable linear and nonlinear solvers with efficient scaling on parallel computers.
Two Firedrake features of particular importance to the current work will be described here: the automatic differentiation capability, and the support for time parallel communications.

\subsection{Automatic differentiation and adjoint models}\label{sec:firedrake:adjoint}

\begin{listing}[t]
\begin{minipage}{0.94\linewidth}
\begin{lstlisting}[style=framed, frame=single, language={[firedrake]python}]
from firedrake import *
from firedrake.adjoint import *

mesh = UnitSquareMesh(40, 40)

# Specify the weak form
V = FunctionSpace(mesh, "CG", 1)
u, v = Function(V), TestFunction(V)
m = Function(V)

a = u*v*dx + inner(grad(u), grad(v))*dx
L = m*v*dx

# Specify the target solution
x, y = SpatialCoordinate(mesh)
u_target = Function(V).interpolate(
    cos(2*pi*x)*cos(2*pi*y))
m.assign(0.)

params = {
    'ksp_type': 'cg',
    'pc_type': 'hypre',
}

# Record the forward model
continue_annotation()
solve(a-L==0, u, solver_parameters=params)
J = assemble((u-u_target)**2*dx)
Jhat = ReducedFunctional(J, Control(m))
pause_annotation()

# Specify and solve the optimiser
tao_params = {
    'tao_monitor': None,
    'tao_type': 'lmvm',
    'tao_gttol': 1e-2,
}

tao = TAOSolver(MinimizationProblem(Jhat),
                parameters=tao_params)
mopt = tao.solve()
\end{lstlisting}
\caption{
   Firedrake code for an inverse problem to find the forcing function \(m\) of the Poisson equation \(\nabla^2 u=m\) that gives a target solution \(u=u_{\textrm{target}}\).
   Firedrake and UFL components are highlighted in blue, and Pyadjoint components in orange.
}
\label{lst:adjoint}
\vspace{-2em}
\end{minipage}
\end{listing}

Firedrake \citep{FiredrakeUserManual} uses the Pyadjoint automatic differentiation library \citep{mitusch_dolfin-adjoint_2019} to enable users to easily solve PDE-constrained optimisation problems.
Listing~\ref{lst:adjoint} details the code required to solve the following inverse problem: find the forcing function \(m\) required to produce a particular solution \(u_{\mathrm{target}}\) to the Poisson problem.
More precisely, we seek the function \(m\in V\) which minimises the objective functional
\begin{equation}\label{fd:eq:objective}
   \hat{J}(m) = \int(u(m)-u_{\mathrm{target}})^{2}\mathrm{d}x
\end{equation}
where \(u(m)\in V\) is the solution of the Poisson equation
\begin{equation}\label{fd:eq:poisson}
   \int uv\mathrm{d}x + \int\nabla u\cdot\nabla v\mathrm{d}x = \int mv\mathrm{d}x \quad \forall v\in V.
\end{equation}
The variational problem for Eq. (\ref{fd:eq:poisson}) on the unit square is set up using UFL on lines 4--12 of Listing \ref{lst:adjoint}, with \(V\) being the space of piecewise linear functions.
The target solution on lines 16--17 is obtained when \(m=(1+8\pi^{2})u_{\mathrm{target}}\), and we use a zero initial guess for \(m\).
Next, on lines 20-21 we specify the parameters for PETSc to solve the assembled linear system for Eq. (\ref{fd:eq:poisson}); we apply the conjugate gradient Krylov (\texttt{ksp}) method using the Hypre multigrid library \citep{hypre} as a preconditioner (\texttt{pc}).

Now we have everything ready to run the forward model, i.e. to calculate \(\hat{J}\) for a given \(m\).
Between the calls to \texttt{continue\_annotation} and \texttt{pause\_annotation}, Firedrake records all operations as Pyadjoint \texttt{Block} objects on Pyadjoint's \texttt{Tape}.
Crucially, operations are taped at a high level of abstraction, e.g. ``solve this variational problem'', rather than at the level of individual floating point or array operations.
For this problem there are two blocks: one for solving the variational problem \(a=L\) for \(u\) on line 27, and one for calculating the norm of the misfit on line 28.
Inside the call to \texttt{solve}, the linear system is obtained by using UFL to symbolically differentiate \texttt{a-L} wrt. \texttt{u}.
C code is automatically generated from the symbolic expression to assemble the linear system, which is then passed to PETSc to solve with the provided parameters.

A Pyadjoint \texttt{ReducedFunctional} is created on line 27 
which provides methods to re-evaluate \(\hat{J}(m)\) at different values of the control \(m\) and to calculate the derivative \(\nabla\hat{J}(m)\) via the adjoint method, as well as to evaluate the actions of the tangent linear and Hessian models.
To re-evaluate \(\hat{J}\) Pyadjoint loops through the blocks on the tape, which each provide a callback to re-evaluate themselves with a new input.
For example the solve \texttt{Block} simply re-solves the UFL expression for the variational form with a new forcing term.
Calculating the derivative \(\nabla\hat{J}(m)\) requires looping through the tape in reverse order, with each block providing a callback for its own derivative.
For example, the derivative of the solve block requires solving the adjoint of the variational form.
To do this, Firedrake uses UFL to symbolically calculate the adjoint form, then sets up a PETSc solver for the adjoint form exactly as for the forward problem but with the adjoint input as the right hand side.
Equivalent callbacks are provided for the TLM and Hessian actions of each block.

Lastly, lines 31--37 set up and solve the optimisation problem with TAO, PETSc's optimisation library, configured again with a dictionary of parameters.
Here we use the Limited Memory Variable Metric method (usually called BFGS \citep{nocedalNumericalOptimization2006a}) with a tolerance of two orders of magnitude drop of the gradient, and printing (monitoring) information at each iteration.
BFGS requires re-evaluation of \(\hat{J}(m)\) and its derivative \(\nabla\hat{J}(m)\) with new values of \(m\), for which Pyadjoint provides callbacks.

This careful coordination of UFL, Pyadjoint, PETSc, and TAO, means firstly that the user only has to define the forward model to get access to the adjoint models essentially ``for free'', and secondly that Firedrake can reuse many of the same software components for both the forward and adjoint models.
A natural consequence is of course that modifying the model or the solution methods is very straightforward.
To change the forward model, e.g. adding a diffusion coefficient or using a different norm to calculate the objective functional requires minimal changes to the code (line 11 or line 26 respectively), and the adjoint callback provided to TAO would automatically change accordingly.
To use a Newton method rather than BFGS would only require changing the TAO parameters, and the TLM and Hessian models would immediately be available.
Finally, to use spatial parallelism only requires running the script with multiple MPI ranks, with no changes to the code necessary.
Firedrake's ability to provide the adjoint for arbitrarily complex models has been used to solve inverse problems in a range of fields, including geophysics \citep{ghelichkhan_automatic_2024}, shape optimisation \citep{paganini_fireshape_2020}, and FWI \citep{roberts_spyro_2022}, among others, as well as enabling tight integration with machine learning frameworks \citep{bouziani_differentiable_2024,farsi_missing_2026}.

\subsection{Time parallelism}\label{sec:firedrake:ensemble}

\begin{figure}
   \centering
   \includegraphics[width=8cm]{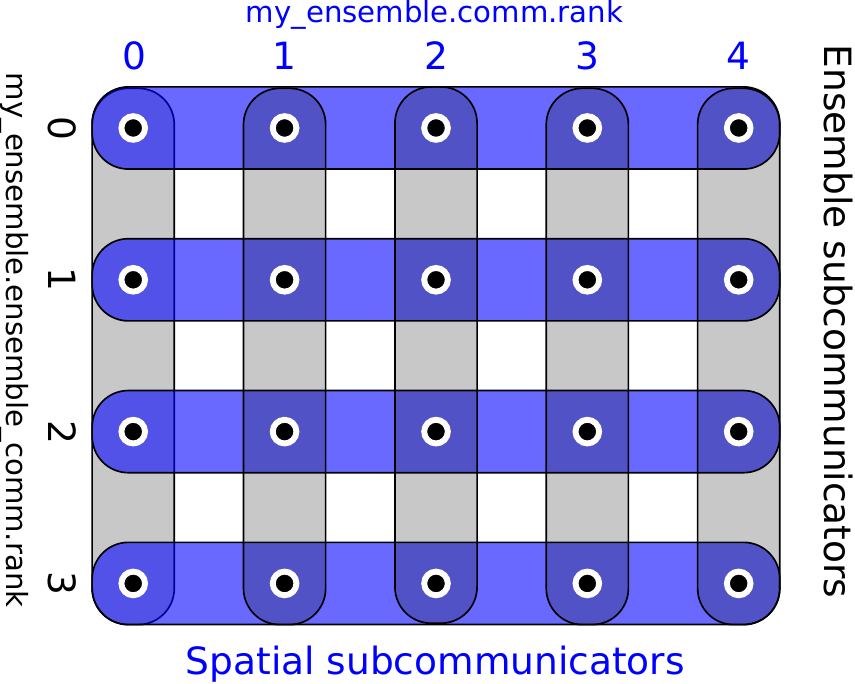}
   \caption{
      The layout for the MPI communicators in Firedrake's \texttt{Ensemble} class.
      Each black circle represents a single MPI rank.
      The blue horizontal boxes are spatial communicators on which a spatial mesh is constructed (the \texttt{Ensemble.comm} attribute).
      Each observation stage is collective over a single spatial communicator, and each spatial communicator may hold multiple stages.
      The grey vertical boxes are temporal communicators which the link the same part of the mesh on all spatial communicators (the \texttt{Ensemble.ensemble\_comm} attribute).
      These communicators are used for communications in time, for example the forward-in-time halos required for the all-at-once system \(\mathbf{L}\), Eq.~(\ref{methods:eq:L-block}), or the backward-in-time halos required for its adjoint \(\mathbf{L}^{T}\). 
      Figure reproduced from \citet{FiredrakeUserManual}.
   }
   \label{fig:ensemble-diagram}
\end{figure}

The second existing Firedrake feature used, and extended, in this work is the support for time-parallelism, where copies of the same spatial domain are created on different groups of processors so that operations on multiple timesteps can be executed simultaneously.
This is supported via the \texttt{Ensemble} communicator object, which uses mpi4py \citep{dalcin_mpi_2005,dalcin_mpi4py_2021}.
It is important to note immediately that the \texttt{Ensemble} object does not imply the use of ensemble data assimilation methods, although its use for such methods was the reason for its naming.

The \texttt{Ensemble} is based on the Message Passing Interface (MPI) parallel model.
In MPI, each processor is called a \emph{rank}, and groups of ranks are called \emph{communicators} or \emph{comms}.
Ranks within the same communicator can communicate data (called messages) to each other.
If an operation requires communication between all ranks in a communicator then it is called a \emph{collective} operation.
Examples of collective operations include \emph{reductions} that calculate the sum of values from all ranks in the communicator, and linear solves where the data is parallelised over the communicator and halo swaps are required.

The \texttt{Ensemble} splits a \emph{global} MPI communicator (usually \texttt{COMM\_WORLD}) into a Cartesian product of smaller communicators, shown in Fig. \ref{fig:ensemble-diagram}, where each black circle represents one rank.
The blue communicators are referred to as \emph{spatial communicators} because a copy of the spatial domain is created over each one.
Communications for spatial parallelism, e.g. halo swaps, are sent over these communicators independently of each other.
The grey communicators are referred to as \emph{ensemble communicators}, and connect the same section of the spatial mesh of each spatial communicator.
Communications for temporal parallelism, e.g. ``time-halos'' for the previous or next timesteps are sent over these communicators.

There is already support for a particular kind of adjoint calculation with the \texttt{Ensemble}, in the case where the objective functional \(\hat{J}(m)\) is the sum of many independent terms \(\hat{J}_{i}(m)\) which all depend on the same control \(m\):
\begin{equation}\label{firedrake:eq:ensemble-rf}
   \hat{J}(m) = \sum^{N}_{i=1} \hat{J}_{i}(m),
   \quad
   \nabla\hat{J} = \sum^{N}_{i=1} \nabla\hat{J}_{i}.
\end{equation}
This situation arises for example in full waveform inversion, where each \(\hat{J}_{i}\) corresponds to a different wave source \citep{roberts_spyro_2022}.
The \texttt{EnsembleReducedFunctional} can be used to distribute the terms \(\hat{J}_{i}\) over different spatial comms of the \texttt{Ensemble} so that the forward and adjoint models for different terms can be calculated in parallel.
However, even though the \textit{computations} are distributed, the \textit{control} is not, i.e. \(m\) must be defined in space only (e.g. a \texttt{Function}).
This means that the \texttt{EnsembleReducedFunctional} is not suitable for WC4DVar because the control \(\mathbf{x}\) for the WC4DVar objective functional \(\hat{J}_{w}(\mathbf{x})\) is defined over both space \textit{and} time.

\section{Aims}\label{sec:aims}
Now that we have described the methods with which we are concerned, and the software that are using, we can formulate the aims and state the contributions of the current work more precisely.
In the demo in the previous section, we showed that Firedrake enables the use of adjoint based methods for PDE constrained optimisation, with spatial parallelism, with almost no extra code compared to simply implementing the forward model.
\textit{The overarching aim of the current work is to enable the use of weak constraint 4DVar with almost no extra code compared to simply implementing the forward model.}
We include in this the use of space-time parallelism and suitable preconditioners.
Requiring only the forward model has benefits for both users and developers.
For application scientists who want to try 4DVar, it means that they can use their expertise to design and implement the forward model without the need for (acquiring) additional expertise in adjoint calculations.
For 4DVar method developers, although they will already have expertise in adjoint methods, it massively reduces the overhead of taking a novel method through a hierarchy of increasingly complex models building up to their target application.

In order to break this broad overarching aim into more concrete points, we establish the four steps required to run a 4DVar model.
\begin{enumerate}
   \item[S1.] Specification of the components of \(J_{w}\) (i.e. \(\mathcal{M}\), \(\mathcal{H}\), \(B\), \(Q\), and \(R\)).
   \item[S2.] Construction of \(J_{w}\) and its adjoint, tangent linear, and Hessian models from its components.
   \item[S3.] Specification of the nonlinear and linear solvers (e.g. Newton with the saddle point formulation and \(\mathbf{P}_{U}\)).
   \item[S4.] Construction of the linear solver components from \(J_{w}\) (e.g. \(\tilde{\mathbf{S}}\), \(\mathbf{A}\), \(\mathbf{P}_{U}\) etc).
\end{enumerate}
Application scientists and numerical analysts who are using and developing 4DVar methods are primarily interested in Steps 1 and 3.
On the other hand, Steps 2 and 4 represent the computational implementation of Steps 1 and 3 respectively, and as such are not usually of interest in and of themselves for these audiences.
Additionally, elements of Steps 2 and 4 require different skillsets than those usually required for Steps 1 and 3, for example efficient implementation of space-time parallelism and automatic differentiation.
With these points in mind, we identify three distinct aims for the current work:
\begin{enumerate}
   \item[A1.] Automate the construction of \(J_{w}\) (S2) while allowing the user full control over the specification of its components (S1).
   \item[A2.] Automate the construction of the linear systems from \(J_{w}\) (S4), while allowing the user full control over the specification of the linear solver (S3).
   \item[A3.] Provide performant space-time parallelism without requiring the user to coordinate parallel communications.
\end{enumerate}

We highlight two common types of workflow that these aims enable.
Firstly, automating the construction of \(J_{w}\) (A2) permits users to easily change the solvers while keeping the forward model fixed.
This is important for application scientists who want to find the best method for their model, and for numerical analysts in the early stages of developing a novel method.
Secondly, the combination of A2 and the automated construction of the linear systems (A1) facilitates easy changes to the forward model while keeping the solver fixed.
This is important for numerical analysts who want to try their novel method on a range of applications. 

The major contributions of the current work are therefore twofold.
Firstly, to achieve A1 we introduce a new \texttt{WC4DVarReducedFunctional} class, which constructs the parallel-in-time WC4DVar objective functional \(J_{w}\) and its derivatives, and provides a user-friendly interface for providing the various required components (described in Sec. \ref{sec:software:demo}).
Secondly, to achieve A2, we introduce a collection of PETSc matrix (\texttt{Mat}) and preconditioner (\texttt{PC}) python types in Sec. \ref{sec:software:impl:petsc} which implement the various linear systems described in Sec. \ref{sec:methods:lsq} from the \texttt{WC4DVarReducedFunctional}.
A3 is achieved through the implementation of these contributions, with details of differentiating through the time-parallel communications in Sec. \ref{sec:software:impl:autodiff}.
Other notable contributions made during this work include an interface for autoregressive covariance operators (described in Sec. \ref{sec:covariance}), and extensions to the coupling between pyadjoint and TAO (described in Sec. \ref{sec:software:impl:petsc}).
We highlight that because Firedrake already provides the derivative models of the spatially distributed components of \(J_{w}\) (e.g. \(M\), \(H\)), and atomic communications for time-parallelism, the primary software challenge of A1 is the proper combination of these two features in a user-friendly manner.

\section{Covariance operators}\label{sec:covariance}
In Eq. (\ref{methods:eq:LLSweak}) (resp. Eq. (\ref{methods:eq:LLSstrong})) the contribution of prior, model and observation components are weighted by the  matrices $\mathbf{D}^{-1}$ and $\mathbf{R}^{-1}$ (resp. $B^{-1}$ and $R_i^{-1}$). Recall the  assumption that all variables are distributed about a mean value with errors drawn from a zero-mean Gaussian with prescribed error covariance. 

In this software, users are free to prescribe any covariance matrix. In this section we provide a description of two options that are provided with the current implementation. Firstly, in Sect. \ref{sec:covariance:diagonal} we discuss uncorrelated covariance operators. In Sect. \ref{sec:covariance:diffusion} we present an implementation of a diffusion operator that is commonly-used to represent background covariance matrices in data assimilation. In Sect. \ref{sec:covariance:implementation} we describe implementational considerations that allow for the correct and efficient implementation of covariance operators within a FEM structure. 
 These aspects are likely to be highly relevant when implementing new covariance operators. In particular we note that much of the data assimilation literature considers finite difference methods. Working with FEM has an important impact on the construction of covariance operators. Finally, in Sect. \ref{sec:covariance:mixed} we describe how these covariance operators can be extended to mixed systems while maintaining balance between fields.

\subsection{Uncorrelated covariance operators} \label{sec:covariance:diagonal}
For many observation types, the error, or uncertainty, between two different measurements is uncorrelated. To construct the correlation operator for uncorrelated noise we use the definition of white noise \(w\in V\) in a finite element space from \citet{Hida_Kuo_Potthoff_Streit_1993},
\begin{equation}\label{cov:eq:assembled-white-noise}
	b = \langle w, v\rangle,
	\quad
	b_{i} = \langle w, v_{i}\rangle\sim \mathcal{N}(0,M)
\end{equation}
i.e. the assembled vector of \(w\) integrated against all test functions \(v_{i}\in V\) has a mass matrix correlation. We note that in a finite difference setting, the corresponding correlation matrix is given by the identity matrix. 

Uncorrelated covariance matrices can be obtained from uncorrelated correlation matrices via specification of standard deviations for each variable and the relation $C = \Sigma M\Sigma$. The standard deviation matrix, $\Sigma$, is a diagonal matrix with $\Sigma_{ii}$ representing the standard deviation in the $i$-th variable. 

\subsection{Diffusion based correlated operators} \label{sec:covariance:diffusion}
There are a variety of methods for approximating the error covariance operators, including deterministic methods such as auto-regressive (AR) or exponential kernels, online flow dependent approximations from ensemble methods, or using historical data.
One common approach is to model the action of the error covariance operator as the solution of a diffusion equation \citep{Weaver_Courtier_2001,Mirouze_Weaver_2010,weaverDiffusionEquationIts2013}.
This approach has a number of useful properties.
\begin{itemize}
	\item If the diffusion equation is solved exactly then the kernel of the corresponding covariance operator is a Gaussian whose lengthscale can be specified via the diffusivity coefficient.
	\item Although solving exactly is usually infeasible, it is straightforward to solve approximately with standard discretisation methods (e.g. finite elements), and a timestepping method such as the Backward Euler method.
	\item Efficient, scalable preconditioners such as multigrid are available for the resulting linear systems. 
	\item Unlike spectral filters, this approach can be easily applied to arbitrary unstructured grids \citep{guilletModellingSpatiallyCorrelated2019}.
	\item Spatially varying and anisotropic covariance operators can be accounted for with spatially varying or tensor-valued diffusion coefficients respectively.
\end{itemize}
The continuous diffusion equation with constant diffusion coefficient \(\kappa\) acting on an error \(e(x, t)\) is
\begin{equation}\label{cov:eq:diff-continuous}
	\partial_{t}e + \kappa\nabla^{2}e = 0,
	\quad
	e(x, 0) = e_{0}.
\end{equation}
The solution \(e(x, \hat{t})\) of Eq.~(\ref{cov:eq:diff-continuous}) at time \(\hat{t}\) is the convolution of the initial condition \(e_{0}\) with a Gaussian kernel \citep{Weaver_Courtier_2001}.\footnote{The convolution solution requires an infinite domain. For finite domains appropriate boundary conditions must be specified, see \citet{Mirouze_Weaver_2010} Section 4 for further discussion.}
Solving Eq.~(\ref{cov:eq:diff-continuous}) exactly is generally impractical, so instead we can discretise in space and time using finite elements and \(m\) steps of Backward Euler respectively.
This leads to the following expression for the discrete correlation operator
\begin{equation}\label{cov:eq:B-action}
	B = \sigma\lambda_{m}\left(K^{-1}M\right)^{m}M^{-1}\lambda_{m}\sigma
\end{equation}
where \(\sigma^{2}\) is the variance, \(\lambda_{m}\) is a normalisation factor, \(M\) is the finite element mass matrix, and \(K\) is an appropriate discretisation of a single Backward Euler step.
For example, for continuous Lagrange elements \(K\) is the assembled matrix corresponding to the standard weak form for the Poisson equation.
\begin{equation}\label{cov:eq:diff-CG}
	K_{ij} = \langle u_{i}, v_{j}\rangle + \langle\kappa\nabla u_{i}, \nabla v_{j}\rangle
\end{equation}
where \(\langle\cdot,\cdot\rangle\) is the inner product on the finite element function space.
Without loss of generality, we have set \(\Delta t=1\) and \(\hat{t}=m\).
For discontinuous finite element spaces a suitable interior penalty method can be used to discretise the Laplacian term.
\citet{Mirouze_Weaver_2010} and \citet{weaverDiffusionEquationIts2013} show that the kernel of Eq.~(\ref{cov:eq:B-action}) with finite \(m\) is actually the \(m\)-th order AR function, which converges to the Gaussian kernel as \(m\to\infty\).
For a given correlation lengthscale \(\ell\) the diffusivity coefficient and the normalisation factor are given in Section 2.4 of \citet{weaverDiffusionEquationIts2013}.
Equation~(\ref{cov:eq:B-action}) emphasises that the covariance operator maps from the dual space to the primal space, i.e. \(B:V^{\star}\to V\).
The action of \(B\) requires \(m\) solves of the discretised diffusion operator \(K\), for which we have efficient scalable solvers such as multigrid.
The right-multiplication with \(M^{-1}\) in Eq.~(\ref{cov:eq:B-action}) is purely notational to make the exponent of the bracketed expression \(m\), and in practice is cancelled with the first \(M\).

\subsection{Implementational concerns}\label{sec:covariance:implementation}

In this section we discuss some aspects related to implementation of covariance operators. We begin by considering how the covariance operators are used within the 4DVar objective function and preconditioners. Within Eq. (\ref{methods:eq:LLSweak}) and Eq. (\ref{methods:eq:LLSstrong}) the inverse covariance operators are used to evaluate the weighted norms. Forward covariance operators (or approximations) form a part of some preconditioners. In the case of the saddle point reformulation, covariance operators are used in Eq. (\ref{methods:eq:saddlepoint}), and inverse covariance matrices (or approximations) appear within preconditioners.  

For uncorrelated preconditioners applying the inverse $C^{-1} = \Sigma^{-1} M^{-1} \Sigma^{-1}$ is affordable, as it requires only the inverse of a diagonal matrix, and the inverse of the mass matrix.

For diffusion based covariances, the inverse of the covariance operator defined in Eq.~(\ref{cov:eq:B-action}) is:
\begin{equation}\label{cov:eq:B-inverse}
	B^{-1} = (\sigma\lambda_{m})^{-1}M\left(M^{-1}K\right)^{m}(\lambda_{m}\sigma)^{-1}.
\end{equation}
This is computationally cheaper to apply than \(B\), requiring only \(m\) matrix actions of \(K\) and \(m-1\) solves of \(M\), which can be effectively preconditioned by its diagonal \citep{Wathen_Rees_2009}.
For more complex operators, an approximate inverse operator could be implemented separately to be used within preconditioners. 

We note that some preconditioners, for example the split preconditioner applied in Eq. (\ref{methods:eq:symmetricL1wc}), require the application of the square root matrix i.e. $G$ such that $C =GG$.
We will now discuss how a not necessarily symmetric decomposition of a covariance matrix can be used to 1) improve the efficiency of evaluation of the weighted norms and 2) generate samples.

Equation~(\ref{cov:eq:B-inverse}) implies an efficient implementation of the weighted norm \(\|x\|_{B^{-1}}=x^{T}B^{-1}x\) in the case that  $m$ is taken to be even \citep{weaverDiffusionEquationIts2013}.
The standard \(L2\) norm is \(\|x\|_{M}=x^{T}Mx\).
Equation~(\ref{cov:eq:B-inverse}) can be rearranged to insert a factor of \(M\) in the middle:
\begin{equation}
	B^{-1} = (\sigma\lambda)^{-1}(KM^{-1})^{m/2}M(M^{-1}K)^{m/2}(\lambda\sigma)^{-1}
\end{equation}
The weighted norm can then be expressed in terms of the standard L2 norm.
\begin{equation}\label{cov:eq:weighted-norm}
	\begin{aligned}
		\hat{B}^{-1/2} & = (M^{-1}K)^{m/2}(\lambda\sigma)^{-1} \\
		B^{-1} & = \hat{B}^{-T/2}M\hat{B}^{-1/2} \\
		\|x\|_{B^{-1}} & = \|\hat{B}^{-1/2}x\|_{M}
	\end{aligned}
\end{equation}
Using Eq.~(\ref{cov:eq:weighted-norm}) requires only \(m/2\) applications of \(M^{-1}K\) followed by a standard L2 norm, instead of a full \(m\) applications for the naive implementation using Eq.~(\ref{cov:eq:B-inverse}).
This formulation also provides an intuitive understanding of the effect of using \(B^{-1}\) to weight the error norms.
The diffusion operator \(K^{-1}\) will smooth the solution, especially high frequency modes with variations below the correlation lengthscale \(\ell\).
Conversely, \(K\) is antidiffusive and will amplify variations, especially the high frequency modes.
These are modes that have a low probability of being drawn from \(\mathcal{N}(0,B)\), so it is only right that they are penalised more heavily in the objective functional. 

The covariance operator can also be used to generate synthetic data for testing a data assimilation implementation. 
When developing new methods it is often useful to test on synthetic data because this allows a greater degree of control than testing immediately on real-world data, and/or because real-world data is not available (e.g. for the Burgers equation).
A common method for generating synthetic data is to run a precursor simulation using a known initial condition \(\hat{x}\) and adding noise at appropriate times.
There are two approaches to introducing error into the synthetic data.
The first approach is to run the ground-truth simulation with different problem parameters than used in the 4DVar model for example a different \(\Delta t\) or layer depth in the semigeostrophic equations \citep{Fisher_Gürol_2017}.
The second approach is to add noise a) to \(\hat{x}\) to generate the background \(x_{b}\) and the ground-truth initial condition \(\hat{x}_{0}\) to emulate uncertainty in previous forecasts, b) to the ground truth state \(\hat{x}_{i}\) at observation times to emulate errors in the PDE model, and c) to the observations \(y_{i}\) to emulate measurement error.
In doing so, the ground-truth data \(\mathbf{\hat{x}}\) is no longer an exact solution to \(\mathcal{M}\), consistent with the weak constraint assumption.
To be consistent with the problem specification, this noise should be drawn from distributions \(\mathcal{N}(0,C)\), where \(C\in\{B,Q_{i},R_{i}\}\) as appropriate.
The process of generating synthetic data in this way is detailed in Alg. \ref{cov:alg:synthetic-data}.
Using a shared initial state \(\hat{x}\) to define \(x_{b}\) and \(\hat{x}_{0}\) means that they will both contain noise even if the input to Alg. \ref{cov:alg:synthetic-data} is specified deterministically.

\begin{algorithm}[t]
\begin{algorithmic}[1]
   \STATE \textbf{input} \(\hat{x}\)
   \STATE \(x_{b} \gets \hat{x} + B^{1/2}z^{V}_{b}\)
   \STATE \(\hat{x}_{0} \gets \hat{x} + B^{1/2}z^{V}_{0}\)
   \STATE \(y_{0} \gets \mathcal{H}(\hat{x}_{0}) + R_{0}^{1/2}z^{U}_{0}\)
   \FOR {\(i=1\) \TO \(i=N\)}
      \STATE \(\hat{x}_{i} \gets \mathcal{M}(\hat{x}_{i-1}) + Q_{i}^{1/2}z^{V}_{i}\)
      \STATE \(y_{i} \gets \mathcal{H}(\hat{x}_{i}) + R_{i}^{1/2}z^{U}_{i}\)
   \ENDFOR
   \RETURN \(x_{b}\), \(\mathbf{\hat{x}}\), \(\mathbf{y}\)
\end{algorithmic}
\caption{Algorithm for generating synthetic data with specified error covariances from a known initial state \(\hat{x}\): ``ground-truth'' trajectory \(\mathbf{\hat{x}}\) with error covariances \(Q_{i}\); observations \(\mathbf{y}\) with error covariances \(R_{i}\); and a background state \(x_{b}\) with error covariance \(B\). The vectors \(z_{j}^{X}\) are a white noise samples in the function space \(X\) for use in Eq. \ref{cov:eq:correlate-noise}.}
\label{cov:alg:synthetic-data}
\end{algorithm}

For any factor \(G\) of a covariance matrix \(B=GG^{T}\), we can generate samples from \(\mathcal{N}(0,B)\) using:
\begin{equation}\label{cov:eq:correlate-noise}
	b = Gz \sim \mathcal{N}\left(0, B\right) \; \text{if} \; z\sim\mathcal{N}\left(0,I\right)
\end{equation}
which can be seen by remembering \(E\left[zz^{T}\right]=I\) and considering:
\begin{equation}
	E\left[bb^{T}\right] = E\left[Gzz^{T}G^{T}\right] = GE\left[zz^{T}\right]G^{T} = B
\end{equation}

Then, from a mass matrix square root \(M=G_{M}G_{M}^{T}\) we can generate \(b\) from Eq.~(\ref{cov:eq:correlate-noise}) and \(w\) by solving Eq.~(\ref{cov:eq:assembled-white-noise}).
\citet{Croci_Giles_Rognes_Farrell_2018} show that a suitable \(G\) can be constructed by first calculating the \(G_{e}\), the element-wise Cholesky factorisation of the element mass matrix \(M_{e}=G_{e}G_{e}^{T}\), and then assembling the local factorisations into the global \(G_{M}\) using the standard finite element assembly procedure.
This approach requires only local computation and communication and, unlike global factorisations, the arithmetic operation count scales linearly with the number of DoFs.
We note that if a factorisation $G_M$ of $M$, is available then $C = \Sigma M \Sigma = (\Sigma G_M)(\Sigma G_M)^T$ is an efficient factorisation of a general uncorrelated covariance matrix.

An efficient matrix factorization \(B=G_{B}G_{B}^{T}\) for the diffusion operator covariance matrix can be calculated for the case where $m$ is even by rewriting Eq.~(\ref{cov:eq:B-action}) as 
\begin{align}
	B & = \sigma\lambda M^{-1}(MK^{-1})^{m/2}M(K^{-1}M)^{m/2}M^{-1} \lambda\sigma \\
	G_{B} & = \sigma\lambda M^{-1}(MK^{-1})^{m/2}M^{1/2} \\
	& = \sigma\lambda M^{-1}(MK^{-1})^{m/2}G_{M}. \label{cov:eq:B-sqrt}
\end{align}
Correlated noise samples can be generated using Eq.~(\ref{cov:eq:B-sqrt}) by first generating an uncorrelated noise sample with \(G_{M}\) and then integrating \(m/2\) steps of the diffusion equation.
We comment that having a factorisation \(B=G_{B}G_{B}^{T}\) is necessary to generate samples from \(B\), but is not necessary for solving the 4DVar system - it simply enables the more efficient implementation of the weighted norms in Eq.~(\ref{cov:eq:weighted-norm}).

Support for covariance operators has been implemented in Firedrake via the \texttt{CovarianceOperatorBase} class, which defines a general interface for covariance operators including applying the action, inverse, weighted norm, and generating correlated noise samples.
The \texttt{AutoregressiveCovariance} provides such an implementation for the diffusion-based covariance operators in Eq.~(\ref{cov:eq:B-action}), but new operators can be defined by users with the provided base class.
Lastly, the covariance operators are required in the matrices \(\mathbf{D}\), \(\mathbf{D}^{-1}\), and \(\mathbf{R}\) described in Sect. \ref{sec:methods:lsq}.
A PETSc matrix can be created from any \texttt{CovarianceOperatorBase} class using the \texttt{CovarianceMat}, where the matrix-vector multiplication can be either the action or inverse of the covariance operator as required.
Conversely, a preconditioner for this matrix type can be created with a \texttt{CovariancePC}, which will automatically apply the opposite operation (e.g. if the Mat applies the inverse of \(B\), then the PC will apply the action).
Use of these types will be demonstrated later in Sect. \ref{sec:software:demo}.

The \texttt{WhiteNoiseGenerator}  generates uncorrelated noise using Eq.~(\ref{cov:eq:assembled-white-noise}) and the element-wise mass matrix factorisation described above.
Two implementations of the element-wise mass matrix are available, the first uses PyOP2 \citep{Rathgeber_Markall_Mitchell_Loriant_Ham_Bertolli_Kelly_2012} to generate C kernels for the element-wise factorisations using LAPACK and the global matrix assembly. The second uses PETSc to calculate calculate the Cholesky factorisation of the block-diagonal mass matrix of a discontinuous superspace of \(V\).

\subsection{Covariance matrices for mixed systems}\label{sec:covariance:mixed}
So far in this section we have considered variables with a single component, meaning that only spatial correlations need to be taken into account.
However, many systems of interest have multiple components, for example atmospheric models include multiple velocity components and thermodynamic variables.
In this case both spatial and multivariate correlations must be taken into account.
For example, a system whose state has two components \(x=(u_{0}, u_{1})\) will in general have a covariance matrix of the form
\begin{equation}\label{cov:eq:general-multivariate-B}
   B =
   \begin{pmatrix}
      B_{00} & B_{01} \\
      B_{10} & B_{11} \\
   \end{pmatrix}
\end{equation}
where \(B_{ij}=B_{ji}\) is the covariance between \(u_{i}\) and \(u_{j}\).
A naive approach would be to assume that the components are only correlated in space and not with each other, i.e. any off-diagonal block (\(B_{ij}, i\neq j\)) is zero, and then each \(B_{ii}\) can be modelled using one of the techniques described above.
However, this assumption is unlikely to be representative of the real physical system.
As described in \citet{bannisterReviewForecastError2008,bannisterReviewForecastError2008a}, a common approach in the atmospheric sciences community is to separate the treatment of the multivariate and spatial correlations.
We give a brief summary of this approach as it may be unfamiliar to readers from other communities, and we will use it for some of the examples in Sect. \ref{sec:examples}.
The interested reader is directed to \citet{bannisterReviewForecastError2008a} for an excellent description of this approach in the context of operational weather forecasting.

The covariance operator is defined using a factorisation
\begin{equation}\label{cov:eq:cvt-factorisation}
   \quad B = K\hat{B}K^{T} = K\hat{B}^{1/2}C\hat{B}^{T/2}K^{T}
\end{equation}
which defines the following sequence of variables:
\begin{equation}\label{cov:eq:cvt-transforms}
\begin{aligned}
   x       \sim \mathcal{N}(0,B),       & \\
   \hat{z} \sim \mathcal{N}(0,\hat{B}), & \quad x=K\hat{z}, \\
   z       \sim \mathcal{N}(0,C),       & \quad z=\hat{B}^{1/2}\hat{z}.
\end{aligned}
\end{equation}
The components of the state \(x\) have both spatial and multivariate correlations.
A suitable choice of \(K\) will account for the multivariate correlations such that the components of \(\hat{z}=(\hat{z}_{0},\hat{z}_{1},\dots)\) will be correlated in space but not with each other, meaning that \(\hat{B}\) is block diagonal.
Note that \(\hat{z}\) does not necessarily have the same number of components as \(x\).
In turn, a suitable choice of \(K\) will account for the spatial correlations in \(\hat{z}\) such that \(z\) is approximately uncorrelated and \(C\approx I\).
While spatial covariance operators may often be reused between different application areas, \(K\) will contain significant knowledge of the physical system.
For example, in atmospheric modelling \(K\) will usually encode balance relations between pressure and Coriolis forces either through analytical relations or regression operators.
Then \(\hat{z}\) has components corresponding to the balanced and unbalanced parts of the flow, which can each be assigned different variances and spatial covariance operators in \(\hat{B}\).

The factorisation of \(B\) is often used as a ``control variable transform'' so that the optimisation is carried out with \(\hat{z}\) or \(z\) as the control.
When solving the least-squares problem for the strong-constraint formulation Eq.~(\ref{methods:eq:primalstrong}), such a transform is equivalent to symmetrically preconditioning the Hessian by \(K\) or \(K\hat{B}^{1/2}\) respectively.
For the weak-constraint formulation a similar transformation can be used assuming that \(B\) and all \(Q_{i}\), and hence \(\mathbf{D}\), have the factored form of Eq.~(\ref{cov:eq:cvt-factorisation}).

In fdvar \(\hat{B}\) is constructed from a list of covariance operators for each component with the \texttt{MixedCovarianceOperator} class, and users can define a transform \(K\) suitable for their own application area using standard Firedrake functionality.
These operations can be composed using the \texttt{CovarianceOperatorBase} class to define general multivariate covariance operators.

\section{Observation operators}\label{sec:obsoperators}

The observation operator \(\mathcal{H}\) plays an important role in mapping from state space $V$ to observation space $U_i$ at each observation time.
We distinguish between two classes of observation operators: direct observations and indirect operators.

\subsection{Direct observations}
Direct observations correspond to measurements of the state quantity.
These can be either point observations at mesh locations or between mesh points.
In both cases the corresponding operator is a linear interpolation.
We implement any direct observations with point measurements using Firedrake's \texttt{VertexOnlyMesh}, which interpolates from \(V\) onto a point cloud \citep{nixon-hill_consistent_2024}.
This interpolation is a symbolic operation so can be differentiated through to calculate the tangent linear and adjoint models.
The observations are numerically evaluated as required by ``assembling'' the symbolic interpolation.

\subsection{Indirect observations}
In practical applications it is common that the measured variables have a more complex relationship with the state variables.
In many cases this leads to an observation operator that is nonlinear.
For example, satellite observations  measure brightness temperature, which relates to `physical' variables such as temperature and humidity via the radiative transfer equations \citep{saundersUpdateRTTOVFast2018}.
In this case, there are similar challenges related to obtaining tangent linear and adjoint models that occur for a non-linear model $\mathcal{M}$.
Using the same automatic differentiation tools and software infrastructure as for the model terms, we can readily obtain the necessary linearised operators to account for more complex observation operators.
For example, the mesh-to-mesh interpolation capability of Firedrake can be used to take observations along an smaller and/or lower dimensional mesh embedded in the original mesh.
Further transforms can then be applied on the embedded mesh, for example integration over the mesh or additional solves for observed quantities.

Therefore users are free to prescribe direct or indirect observation operators with no change in the remaining workflow. 
We provide an example of a nonlinear observation operator in Sect. \ref{sec:examples:NLSWE}.

\section{4DVar methods in Firedrake}\label{sec:software}
In this section  we walk through code examples for a typical script in Sec. \ref{sec:software:demo}, highlighting the user \textit{interface} and how code components correspond to the mathematical objects in Sec. \ref{sec:methods}.
We will present the numerical results for this example later in Sect. \ref{sec:examples:Burgers results}.
We then describe the novel software contributions and \textit{implementation} in Sec. \ref{sec:software:impl}, highlighting aspects which facilitate the user friendly interface.

\subsection{Burgers' equation demo}\label{sec:software:demo}

To demonstrate the new functionality in action we will walk through code snippets from a typical workflow solving a WC4DVar problem.
The example we use is the Burgers equation in 1D with direct observations at random locations.

\subsubsection{Specifying the forward model}\label{sec:software:demo:step-1}

\begin{listing}[t]
\begin{minipage}{0.94\linewidth}
\begin{lstlisting}[style=framed, frame=single, language={[firedrake]python}]
# Space-time parallelism
ensemble = Ensemble(COMM_WORLD, 1)
mesh = PeriodicUnitIntervalMesh(
   nx, comm=ensemble.comm)
V = FunctionSpace(mesh, "CG", 1)
W = EnsembleFunctionSpace(
   [V]*nlocal_spaces, ensemble)
...
# Burgers equation propagator
u, uprev = Function(V), Function(V)
uh = (u + uprev)/2
F = (inner(u - uprev, v)/dt
     + inner(dot(uh, grad(uh)), v)
     + inner(nu*grad(uh), grad(v)))*dx

def M(x):
   u.assign(x)
   for _ in range(nt):
      uprev.assign(u)
      solve(F==0, u)
   return u.copy(deepcopy=True)
...
# Point observations
point_locations = np.random.rand(20, 1)
point_cloud = VertexOnlyMesh(
   mesh, vertexcoords=point_locations)
U = FunctionSpace(point_cloud, "DG", 0)

def H(x):
   return assemble(interpolate(x, U))
...
# Covariance operators
B = AutoregressiveCovariance(
   V, L_B, sigma_B, m=2)
Q = AutoregressiveCovariance(
   V, L_Q, sigma_Q, m=2)
R = AutoregressiveCovariance(
   U, 0, sigma_R, m=0)
...
# Synthetic "ground-truth" observations
x0 = Function(V).assign(xhat+B.sample())
y = [Function(U).assign(H(x0)+R.sample())]
# state at time i and i-1
xi, xim1 = Function(V), Function(V)

with ensemble.sequential(state=x0) as ctx:
   xi.assign(ctx.state)
   for _ in range(nlocal_stages):
      xim1.assign(xi)
      xi.assign(M(xim1) + Q.sample())
      y.append(H(xi) + R.sample())
   ctx.state.assign(xi)
\end{lstlisting}
\caption{Sample code to specify the forward model for the Burgers equation with implicit midpoint timestepping, direct point observations, and reference data generated from a randomly perturbed forward model. (Step 1).}
\label{lst:step-1}
\vspace{-2em}
\end{minipage}
\end{listing}

Code samples for the major steps in specifying the forward model (i.e. Step 1) are shown in Listing \ref{lst:step-1}.

\textbf{Time parallelism.}
The first step in specifying the forward model (Step 1) is to set up the space-time parallelism, on lines 1-7.
An \texttt{Ensemble} is created from a global communicator, here just \texttt{COMM\_WORLD}, and the number of ranks per spatial comm.
For the 1D problem here we just use 1 rank in space i.e. no spatial parallelism, so the number of spatial comms will be the number of ranks in \texttt{COMM\_WORLD}.
The mesh is created on the spatial comm, and the function space for a single timestep \(V\) is created on the mesh.
For the function space for the entire timeseries \(W\) we use the new \texttt{EnsembleFunctionSpace} class, which represents a mixed function space where the components are distributed over an \texttt{Ensemble}.
This is created by passing the components on the local spatial comm, which for WC4DVar will usually be the \(N_{t}/P_{t}\) copies of \(V\).

\textbf{Propagator.}
On lines 10-22 we define the Burgers equation propagator \(\mathcal{M}\) as \(n_{t}\) steps of the implicit midpoint rule.
The following weak form for the solution \(u \in V\) of the Burgers equation is used on lines 11-15 to define the UFL expression for a single timestep.
\begin{equation}
    F = \int\left(\langle\partial_{t}u + u\cdot\nabla u,\,v\rangle + \langle\nu\nabla u, \nabla v\rangle\right)dx \;\forall v \in V
\end{equation}
We use the default solver parameters for \texttt{solve} for brevity.

\textbf{Observation operator.}
Next we specify the observation operator \(\mathcal{H}\) on lines 25-32.
The observation space \(U\) is discontinuous scalar values (DG0) on a \texttt{VertexOnlyMesh}.
Point evaluation is simply an interpolation of a state \(x\) into \(U\).
We point out that we can interpolate expressions, for example \texttt{interpolate(x*x/2, U)} would take observations of the energy.

\textbf{Covariances.}
For the error covariance operators on lines 35-41 we use the new \texttt{AutoregressiveCovariance} described in Sect. \ref{sec:covariance} with \(Q_{i}=Q\) and \(R_{i}=R\) for simplicity.
Setting \(m=0\) for \(R\) produces uncorrelated covariances.
Note that we could pass parameters for the implicit solves required for the action of \(B\) and \(Q\), for example a multigrid method, but here we just use the default LU solver.

\textbf{Observations.}
The final component of \(\hat{J}_{w}\) to specify is the observations \(y_{i}\in U\), for which we use Alg. \ref{cov:alg:synthetic-data} to generate synthetic data.
The ``ground-truth'' initial condition is created by perturbing some known state \(\hat{x}\) with noise with covariance \(B\), and the observation is perturbed with noise with covariance \(R\).
The code within the new \texttt{Ensemble.sequential} context manager is executed first on spatial comm 0, where any kwarg is made available on the \texttt{ctx} object.
Following Alg. \ref{cov:alg:synthetic-data}, for each local observation stage we propogate \(\mathcal{M}(x_{i-1})\), perturb with noise from \(Q\), observe with \(\mathcal{H}\), and finally perturb with noise from \(R\).
After the last local stage we store \(x_{i}\) back on the \texttt{ctx}.
The \texttt{Ensemble} then sends everything on \texttt{ctx} on to spatial comm 1 and executes the code block within the \texttt{sequential} context manager - i.e. on spatial comm 1 the \texttt{ctx.state} on line 48 is the value saved on line 54 by spatial comm 0.
This means that the user only has to implement the operations on a single observation stage at a time, without having to explicitly interact with the time parallelism.

\subsubsection{Constructing the 4DVar problem}\label{sec:software:demo:step-2}

\begin{listing}[t]
\begin{minipage}{0.94\linewidth}
\begin{lstlisting}[style=framed, frame=single, language={[firedrake]python}]
# Create the collective Jw object
x = EnsembleFunction(W)
xb = Function(V).assign(xhat + B.sample())

initial_obs_error = lambda x: H(x) - y[0]

continue_annotation()
Jhat = WC4DVarReducedFunctional(
   Control(x),
   background=x_b,
   background_covariance=B,
   observation_covariance=R,
   observation_error=initial_obs_error)

# Tape forward model by observation stage
with Jhat.record_stages() as stages:
   for stage, ctx in stages:
      xim1.assign(stage.control)
      xi = M(xim1)

      i = stage.local_index
      obs_error = lambda x: H(x) - y[i]

      stage.set_observation(
         state=xi,
         observation_error=obs_error,
         observation_covariance=R,
         model_covariance=Q)
pause_annotation()
\end{lstlisting}
\caption{
   Sample code to create the WC4DVar reduced functional for the Burgers equation (Step 2).
}
\label{lst:step-2}
\vspace{-2em}
\end{minipage}
\end{listing}

Now that we have created the individual components of the forward model we provide them to fdvar in such a way that the objective functional \(\hat{J}_{w}\) can be built.

This is shown in Listing \ref{lst:step-2} where we use the new \texttt{WC4DVarReducedFunctional} class, which represents \(\hat{J}_{w}\).
We can see that, unlike the standard \texttt{ReducedFunctional}, this new class is created \textit{before} the forward model is recorded rather than after.
We will explain and discuss this design choice later.

To instantiate the \texttt{WC4DVarReducedFunctional} the user passes an \texttt{EnsembleFunction} \(\mathbf{x}\in W\) as the control variable, as well as the background and observation components for the initial time (although these arguments are only used on spatial comm 0).
Next, the forward model for each observation stage is taped using the \texttt{record\_stages} context manager.
On each spatial comm this returns an iterator with a \texttt{stage} object representing each local stage.
To record each stage the user first propagates from \(x_{i-1}\), represented by \texttt{stage.control}, to \(x_{i}=\mathcal{M}(x_{i-1})\).
The stage is finalised by calling \texttt{set\_observation} and providing the final state \(x_{i}\), the error covariances, and a callback for the observation misfit \(d_{i}(x)=\mathcal{H}(x)-y_{i}\).

Internally, \texttt{record\_stages} uses the \texttt{Ensemble.sequential} context manager so that the value of \texttt{stage.control} for stage 0 is \(x_{b}\) and the value of \texttt{stage.control} for stage \(i>0\) is \(x_{i-1}\), i.e. the value of \(x_{i}\) at the end of the previous stage.
This means that the initial guess \(\mathbf{x}^{(0)}\) for the optimisation is a consistent timeseries from \(x_{b}\), and again the user has not had to explicitly deal with the time-parallelism.
Additionally, any kwargs passed to \texttt{record\_stages} will be handled similarly to those passed to \texttt{Ensemble.sequential} and will appear on the \texttt{ctx} at each stage.

Once \texttt{record\_stages} is finished the \texttt{WC4DVarReducedFunctional} is fully initialised and Step 2 is complete, achieving aim A1.
Just like the standard \texttt{ReducedFunctional} we can now re-evaluate \(\hat{J}_{w}\) as well as calculate the adjoint, tangent linear, and Hessian models.
Note that, although the forward model is recorded sequentially by running through the observation stages on each spatial comm in turn, once the \texttt{WC4DVarReducedFunctional} is fully initialised the re-evaluation and derivative calculations are carried out fully parallel-in-time.

Comparing the code in Listing \ref{lst:step-2} to code required to just evaluate the forward model on its own (which would be very similar to lines 48-54 of Listing \ref{lst:step-1}) there are only minor differences.
Firstly, rather than actually executing the covariance-weighted norms and the observation operators they are provided as callbacks.
The other difference is using \texttt{record\_stages} rather than \texttt{Ensemble.sequential} but, because the former depends on the latter, they are syntactically very similar.
On the other hand, if a user already had a state \(\mathbf{x}^{(l)}\) and wanted to evaluate \(\hat{J}_{w}(\mathbf{x}^{(l)})\) parallel-in-time, then they would have to manually handle the time-halos for the first/last observation stage on each spatial comm themselves.

\subsubsection{Solving the primal 4DVar problem}\label{sec:software:demo:step-3-primal}

\begin{listing}[t]
\begin{minipage}{0.94\linewidth}
\begin{lstlisting}[style=framed, frame=single, language={[firedrake]python}]
tao_parameters = {
  'tao_gttol': 1e-3,
  'tao_type': 'nls',
  'tao_ls_type': 'unit',
  'tao_nls': {
    'ksp_type': 'cg',
    'ksp_convergence_test': 'skip',
    'ksp_max_it': 6,
    'pc_type': 'python',
    'pc_python_type': 'firedrake.WC4DVarSchurPC',
    'wcschur_l': {
      'pc_type': 'python',
      'pc_python_type': 'firedrake.AllAtOnceRFGaussSeidelPC',
      'pc_aaogs_type': 'identity'
    },
    'wcschur_d': {
      'pc_type': 'python',
      'pc_python_type': 'firedrake.EnsembleBJacobiPC',
      'sub_pc_type': 'python',
      'sub_pc_python_type': 'firedrake.CovariancePC'
    }
  }
}
\end{lstlisting}
\caption{
   Sample TAO parameters to specify the primal formulation of WC4DVar Eq.~(\ref{methods:eq:NEweak}).
   The linear system at each nonlinear iteration is the primal Hessian \(\mathbf{S}\) from Eq.~(\ref{methods:eq:Hessianweak}) preconditioned with \(\mathbf{\tilde{S}}=\mathbf{\tilde{L}}^{T}\mathbf{D}^{-1}\mathbf{\tilde{L}}\) with \(\mathbf{\tilde{L}}=\mathbf{L}(M=I)\).
}
\label{lst:step-3-primal}
\vspace{-2em}
\end{minipage}
\end{listing}

Now that we have constructed an object for \(\hat{J}_{w}\) we move on to Step 3, specifying the linear and nonlinear solvers in the optimiser, and Step 4, constructing the solver and calculating an optimised state.
For the majority of cases Step 3 is achieved entirely through the TAO parameters dictionary.
Once these parameters are specified, Step 4 is identical to the last 3 lines of Listing \ref{lst:adjoint} where a \texttt{TAOSolver} is created from a reduced functional and a set of TAO parameters.
Typical parameters for the primal formulation Eq.~(\ref{methods:eq:NEweak}) are shown in Listing \ref{lst:step-3-primal}.

\textbf{Primal formulation.}
First we select the Newton/linesearch (\texttt{nls}) optimiser with a unit linesearch (i.e. no search) on lines 3-4 of Listing \ref{lst:step-3-primal}.
Pyadjoint's \texttt{TAOSolver} automatically creates a matrix-free representation of \(\mathbf{S}\) from the reduced functional, and the linear solver is configured in the \texttt{tao\_nls} parameters.
The Krylov method (\texttt{ksp\_type}) is set to CG on line 6. On line 7 we  skip any convergence test and force the KSP to always use the maximum 6 iterations on line 8). On line 10 the preconditioner (\texttt{pc}) is the new \texttt{WC4DVarSchurPC} Python preconditioner, which implements \(\mathbf{\tilde{S}}^{-1}\) and sets up subsolvers to (approximately) solve \(\mathbf{\tilde{L}}\), \(\mathbf{\tilde{L}}^{T}\) via the all-at-once preconditioner, and \(\mathbf{\tilde{D}}^{-1}\) as a covariance preconditioner.

\textbf{All-at-once preconditioner.}
The preconditioner \(\mathbf{\tilde{L}}=\mathbf{L}(M=I)\) is block lower triangular, so we use the new \texttt{AllAtOnceRFGaussSeidelPC} on line 13 and specify to approximate \(M\) with \texttt{identity} on line 14.
For symmetry the same solver parameters are used for both \(\mathbf{\tilde{L}}\) and  \(\mathbf{\tilde{L}}^{T}\), specified in \texttt{wcschur\_l} on lines 11-15.

\textbf{Covariance preconditioner.}
Moving on to \(\mathbf{\tilde{D}}^{-1}\), this is a block diagonal matrix where each block is either \(B^{-1}\) or \(Q^{-1}\), therefore we need a block diagonal preconditioner that sets up a subsolver for each block.
This is specified in the \texttt{wcschur\_d} parameters on line 16-21 using the new \texttt{EnsembleBJacobiPC} on line 18 (BJacobi stands for "block Jacobi").
In the \texttt{sub} parameters for the individual blocks we use the new \texttt{CovariancePC} on line 20 to apply the action of \(B\) or \(Q\) (described in Sect. \ref{sec:covariance}).
The action of the diffusion-based covariance operators itself requires several implicit solves, for which we can provide solver parameters when the \texttt{AutoregressiveCovariance} object is created.

In just a few lines of code we have specified the full solver configuration, with all components constructed automatically entirely from the \texttt{WC4DVarReducedFunctional}.

\subsubsection{Solving the saddle point 4DVar problem}\label{sec:software:demo:step-3-saddle}

\begin{listing}[t]
\begin{minipage}{0.94\linewidth}
\begin{lstlisting}[style=framed, frame=single, language={[firedrake]python}]
tao_parameters = {
  'tao_gttol': 1e-3,
  'tao_type': 'nls',
  'tao_ls_type': 'unit',
  'tao_nls': {
    'ksp_type': 'preonly',
    'pc_type': 'python',
    'pc_python_type': 'firedrake.WC4DVarSaddlePC',
    'wcsaddle': {
      'ksp_type': 'gmres',
      'ksp_rtol': 1e-3,
      'pc_type': 'fieldsplit',
      'pc_fieldsplit_type': 'schur',
      'pc_fieldsplit_0_fields': '0,1',
      'pc_fieldsplit_1_fields': '2',
      'pc_fieldsplit_schur_fact_type': 'upper',
      'fieldsplit_0': {
        'pc_type': 'fieldsplit',
        'pc_fieldsplit_type': 'additive',
        'fieldsplit_0': {...},
        'fieldsplit_1': {...},
      },
      'fieldsplit_1': {
        'pc_type': 'python',
        'pc_python_type': 'firedrake.WC4DVarSchurPC',
        'wcschur_l': {...}
        'wcschur_d': {...}
      }
    }
  }
}
\end{lstlisting}
\caption{
   Sample TAO parameters to specify the saddle point formulation of WC4DVar.
   The linear system at each nonlinear iteration is the saddle point system Eq.~(\ref{methods:eq:saddlepoint}), with \(\mathbf{A}\) preconditioned with \(\mathbf{P}_{U}\) Eq.~(\ref{methods:eq:PU}) where the Schur complement is approximated using \(\mathbf{\tilde{S}}=\mathbf{\tilde{L}}^{T}\mathbf{\tilde{D}}^{-1}\mathbf{\tilde{L}}\) with \(\mathbf{\tilde{L}}=\mathbf{L}(M=I)\).
   The parameters for the subsolvers \texttt{wcschur\_l} and \texttt{wcschur\_d} in \(\mathbf{\tilde{S}}\) are identical to those in Listing \ref{lst:step-3-primal}.
}
\label{lst:step-3-saddle}
\vspace{-2em}
\end{minipage}
\end{listing}

We now demonstrate how the solver can be changed just by modifying the parameters dictionary.
The parameters in Listing \ref{lst:step-3-saddle} specify the saddle point formulation Eq.~(\ref{methods:eq:saddlepoint}), with \(\mathbf{A}\) solved using GMRES preconditioned by \(\mathbf{P}_{U}\), where the Schur complement is approximated using the same \(\mathbf{\tilde{S}}\) as used above for the primal formulation.

\textbf{Saddle point formulation.}
First, we use the \texttt{preonly} KSP type on line 6 to entirely replace the approximate solve of \(\mathbf{S}\) with the application of the new \texttt{WC4DVarSaddlePC} preconditioner.
This PC internally sets up a KSP to (approximately) solve \(\mathbf{A}\), and returns the \(\mathbf{\delta x}\) part of the saddle point solution.
In the \texttt{wcsaddle} parameters we configure the KSP for \(\mathbf{A}\) to use GMRES preconditioned with \(\mathbf{P}_{U}\).

\textbf{Schur factorisation.}
PETSc's ``fieldsplit'' PC provides various block preconditioners, including Schur complement factorisations.
The Schur complement factorisation requires a \(2\times2\) structure, where the first field will be eliminated.
However \(\mathbf{A}\) has three fields (\(\mathbf{\eta}\), \(\mathbf{\lambda}\), and \(\mathbf{\delta x}\)), so we need to tell PETSc how to reinterpret the \(3\times3\) matrix \(\mathbf{A}\) as the \(2\times2\) matrix \(\mathbf{\hat{A}}\) so that it can construct \(\mathbf{P}_{U}\):
\begin{align}\label{eq:2x2-saddle-point}
   \mathbf{\hat{A}} =
   \begin{pmatrix}
      \mathbf{\hat{A}}_{00} & \mathbf{\hat{A}}_{01} \\
      \mathbf{\hat{A}}_{10} & \\
   \end{pmatrix},
   \quad
   \mathbf{P}_{U} =
   \begin{pmatrix}
      \mathbf{\hat{A}}_{00} & \mathbf{\hat{A}}_{01} \\
                            & \mathbf{\tilde{S}} \\
   \end{pmatrix},
   \\
   \mathbf{\hat{A}}_{00} =
   \begin{pmatrix}
      \mathbf{D} & \\
      & \mathbf{R} \\
   \end{pmatrix},
   \quad
   \mathbf{\hat{A}}_{01} =
   \begin{pmatrix}
      \mathbf{L} \\
      \mathbf{H} \\
   \end{pmatrix},
   \quad
   \mathbf{\hat{A}}_{10} =
   \begin{pmatrix}
      \mathbf{L}^{T} & \mathbf{H}^{T}
   \end{pmatrix}.
\end{align}
This is achieved on lines 14-15 where we tell PETSc firstly to eliminate fields 0 and 1 of \(\mathbf{A}\) (\(\mathbf{\eta}\) and \(\mathbf{\lambda}\)), and secondly that we will be building the Schur complement on field 2 of \(\mathbf{A}\) (\(\mathbf{\delta x}\)). 
The \texttt{upper} Schur factorisation type corresponds to \(\mathbf{P}_{U}\).

\textbf{Covariance preconditioner.}
The parameters for solving \(\mathbf{\hat{A}}_{00}\) in the top left block of \(\mathbf{P}_{U}\) are specified in \texttt{fieldsplit\_0} starting on line 17.
Because \(\mathbf{\hat{A}}_{00}\) is block diagonal we can use the \texttt{additive} version of fieldsplit PC to solve for each field independently.\footnote{
   PETSc distinguishes between two types of block preconditioners.
   ``Block Jacobi'' preconditioners allow blocks to exist on different subcommunicators, for example the blocks \(R_{0},R_{1},\dots\) of \(\mathbf{R}\) may be on different spatial communicators.
   On the other hand, ``fieldsplit'' preconditioners require each block (i.e. each field) to exist over the entire global communicator.
}
For each of these fields we can use the same parameters as we used for \(\mathbf{D}^{-1}\) in the \texttt{wcschur\_d} parameters in Listing \ref{lst:step-3-primal}, however this time the \texttt{CovariancePC} will automatically apply the inverse rather than the action of each covariance operator.

\textbf{Schur complement.}
The parameters for the Schur complement \(\mathbf{\tilde{S}}\) in the bottom right block of \(\mathbf{P}_{U}\) are specified in \texttt{fieldsplit\_1} on line 23.
Because the Schur complement of \(\mathbf{A}\) is just the primal Hessian \(\mathbf{S}\) we can use the \texttt{WC4DVarSchurPC} preconditioner again, with the same parameters for \(\mathbf{L}\) and \(\mathbf{D}^{-1}\) as in Listing \ref{lst:step-3-primal}.

\subsection{Software implementation}\label{sec:software:impl}

\subsubsection{Automatic differentiation with nested parallelism}\label{sec:software:impl:autodiff}

\begin{table*}[t]
\caption{
Mappings for fundamental operations on types distributed over an ensemble communicator.
\(\mathbf{x}\in\mathbf{X}\) and \(\mathbf{y}\in\mathbf{Y}\) are objects with \(N+1\) components \(x_{i}\in X_{i}\) and \(y_{i}\in Y_{i}\), for example \texttt{EnsembleFunction}, \texttt{EnsembleCofunction}, or \texttt{EnsembleAdjVec}, whereas \(\hat{x}\in \hat{X}\) and \(\hat{y}\in \hat{Y}\) are objects with a single component, e.g \texttt{Function}, \texttt{Cofunction}, or \texttt{AdjFloat}.
}
\label{tab:ensemble-mappings}
\renewcommand{\arraystretch}{1.25}
\begin{tabular}{clll}
\tophline
Operation  & Mapping & Definition & Requirements \\
\middlehline
Transform & \(\mathcal{T}:\mathbf{X}\to\mathbf{Y}\) & \(y_{i}=\mathcal{T}_{i}(x_{i})\;\forall i\) & \(\mathcal{T}_{i}:X_{i}\to Y_{i}\) \\ \hline
Broadcast & \(\mathcal{B}:\hat{X}\to\mathbf{X}\) & \(x_{i}=\hat{x}\;\forall i\) & \(\mathbf{X}=\hat{X}^{N}\) \\ \hline
Reduce & \(\mathcal{R}:\mathbf{X}\to \hat{X}\) & \(\hat{x}=\sum x_{i}\) & \(\mathbf{X}=\hat{X}^{N}\) \\ \hline
Shift & \(\mathcal{S}:\mathbf{X}\to \mathbf{\hat{X}}\) & \(\hat{x}_{0}=0,\; \hat{x}_{i}=x_{i-1}\;\forall i>0\) & \(\hat{X}_{i}=X_{i-1}\;\forall i>0\) \\
\bottomhline
\end{tabular}
\vspace{2em}
\caption{
Operations in the WC4DVar objective \(\mathcal{J}_{w}\) expressed as compositions of the mappings in Table \ref{tab:ensemble-mappings}.
\(\|\mathbf{x}\|_{\mathbf{D^{-1}}}\) and \(\|\mathbf{x}\|_{\mathbf{R^{-1}}}\) define the weighted space-time norms, \(\mathcal{G}(\mathbf{x})=\mathbf{d}\) evaluates the observation misfits, and \(\mathcal{L}(\mathbf{x})=\mathbf{b}\) evaluates the model misfits.
\(\mathcal{I}(x)=x\) is the identity.
}
\label{tab:wc4dvar-mappings}
\renewcommand{\arraystretch}{1.4}
\begin{tabular}{clll}
\tophline
Operation  & Composition & Derivative & Definitions \\
\middlehline
\(\|\mathbf{x}\|_{\mathbf{R}^{-1}}\)
   & \(\mathcal{R}\circ\mathcal{T}_{\mathbf{R}}\)
   & \(\nabla\mathcal{T}_{\mathbf{R}}^{\star}\circ\nabla\mathcal{R}^{\star}\)
   & \(\mathcal{T}_{\mathbf{R},i}(x)=\|x\|_{R_{i}^{-1}}\;\forall i\) \\ \hline
\(\|\mathbf{x}\|_{\mathbf{D}^{-1}}\)
   & \(\mathcal{R}\circ\mathcal{T}_{\mathbf{D}}\)
   & \(\nabla\mathcal{T}_{\mathbf{D}}^{\star}\circ\nabla\mathcal{R}^{\star}\)
   & \(\mathcal{T}_{\mathbf{D},0}(x)=\|x\|_{B^{-1}}, \;\; \mathcal{T}_{\mathbf{D},i}(x)=\|x\|_{Q_{i}^{-1}}\;\forall i>0\) \\ \hline
\(\mathcal{G}(\mathbf{x})\)
   & \(\mathcal{T}_{\mathcal{G}}\)
   & \(\nabla\mathcal{T}_{\mathcal{G}}^{\star}\)
   & \(\mathcal{T}_{\mathcal{G},i}(x)=\mathcal{H}_{i}(x) - y_{i}\; \forall i\) \\ \hline
\(\mathcal{L}(\mathbf{x})\)
   & \((\mathcal{I} - \mathcal{T}_{\mathcal{M}}\circ\mathcal{S})\)
   & \((\mathcal{I} - \nabla\mathcal{S}^{\star}\circ\nabla\mathcal{T}_{\mathcal{M}}^{\star})\)
   & \(\mathcal{T}_{\mathcal{M},0}(x)=x_{b} \;\; \mathcal{T}_{\mathcal{M},i}(x)=\mathcal{M}_{i}(x)\;\forall i>0\) \\
\bottomhline
\end{tabular}
\end{table*}

In this section we describe how we build on Firedrake's ability to automatically differentiate operations over objects defined in space (e.g. a \texttt{Function}) to re-evaluate the forward and adjoint models of operations objects defined over both space and time.

We refer to \(\mathbf{x}\in\mathbf{X}=X_{0}\times X_{1}\times\dots\) and \(\mathbf{y}\in\mathbf{Y}=Y_{0}\times Y_{1}\times\dots\) as ``ensemble objects'' in ``ensemble function spaces'' which have multiple components,\footnote{
   We reinforce again that this nomenclature does not imply ensemble data assimilation methods, we just aim to be consistent with naming of the \texttt{Ensemble} class.}
(for example \(\mathbf{V}\) and \(\mathbf{U}\)), whereas \(\hat{x}\in\hat{X}\) and \(\hat{y}\in\hat{Y}\) are objects with a single component (for example \(V\) and \(U\)).
In practice a single component \(\hat{x}\in\hat{X}\) is collective only in space i.e. is distributed over a single spatial comm.
On the other hand, an ensemble object \(\mathbf{x}\in\mathbf{X}\) is collective in both space and time i.e. is distributed over the entire \texttt{Ensemble}, with each spatial comm owning one or more components.
The new \texttt{EnsembleFunction} class encodes these collective semantics over the space-time domain.

In Table \ref{tab:ensemble-mappings} we identify four fundamental mappings necessary to describe the time-parallel operations in WC4DVar: Transform \(\mathcal{T}\), Reduce \(\mathcal{R}\), Broadcast \(\mathcal{B}\), and Shift \(\mathcal{S}\).
Mappings can be composed, for example a transform-reduce operation is \((\mathcal{R}\circ\mathcal{T}):(\mathbf{Y}\to\hat{Y})\circ(\mathbf{X}\to\mathbf{Y})\), or equivalently \((\mathcal{R}\circ\mathcal{T}):\mathbf{X}\to\hat{Y}\).
Straightforward application of the chain rule allows us to define the tangent linear and adjoint derivative of a composition of functions as 
\(\nabla(\mathcal{R}\circ\mathcal{T})=\nabla\mathcal{R}\circ\nabla\mathcal{T}\) and  \(\nabla(\mathcal{R}\circ\mathcal{T})^{\star}=\nabla\mathcal{T}^{\star}\circ\nabla\mathcal{R}^{\star}\) respectively.
This is simply a more formal statement of the process described in Sect. \ref{sec:firedrake:adjoint} where Pyadjoint calculates forward or adjoint derivatives by running forward or backwards through the tape respectively.
We note that the \texttt{EnsembleReducedFunctional} described in Sect. \ref{sec:firedrake:ensemble} implements \(\mathcal{R}\circ\mathcal{T}\circ\mathcal{B}:\hat{X}\to\hat{Y}\).

Table \ref{tab:wc4dvar-mappings} shows how each part of the WC4DVar objective \(\mathcal{J}_{w}\) can be expressed in terms of Transform, Reduce, and Shift (the inclusion of Broadcast in Table \ref{tab:ensemble-mappings} will become clear later).
We can express all the operations we need as compositions of four simple mappings, and we can express derivatives of the compositions as compositions of derivatives, so we now just need to define the derivatives of these mappings.

\textbf{Transform}.
This is the simplest mapping as it requires no communication between different elements of \(\mathbf{X}\) (or \(\mathbf{Y}\)).
The tangent linear action and adjoint derivatives are just the corresponding operations of the element transforms applied element-wise, i.e.
\begin{align}\label{eq:transform-derivatives}
   & \nabla\mathcal{T}: \mathbf{X}\to\mathbf{Y},
      \quad
      &&\left(\nabla\mathcal{T}(\delta \mathbf{x})\right)_{i}
         = \nabla\mathcal{T}_{i}(\delta x_{i}) \\
   & \nabla\mathcal{T}^{\star}: \mathbf{Y}^{\star}\to\mathbf{X}^{\star},
      \quad
      &&\left(\nabla\mathcal{T}^{\star}(\delta \mathbf{y}^{\star})\right)_{i}
         = \nabla\mathcal{T}_{i}^{\star}(\delta y^{\star}_{i})
\end{align}
In practice each \(\mathcal{T}_{i}\) is local to a specific spatial comm on the \texttt{Ensemble} and is defined by a \texttt{ReducedFunctional} with local controls (e.g. \texttt{Function}) which will provide the forward and adjoint models when required.

The \(\mathcal{T}_{i}\) for \(\mathcal{T}_{\mathcal{G}}\), \(\mathcal{T}_{\mathbf{D}}\), and \(\mathcal{T}_{\mathbf{R}}\) are created by taping the callbacks for \(\mathcal{H}\) and the covariance operators provided to the \texttt{WC4DVarReducedFunctional}.

\textbf{Reduce \& Broadcast}.
The reduce operation is a sum over all elements of the input.
We can express this as the vector operation
\begin{align}\label{eq:reduce-vec-op}
   \mathcal{R}(\mathbf{x}) = \hat{x} = \sum^{N}_{i=0} x_{i} = \mathbf{e}^{T}\mathbf{x}, \quad \mathbf{e}^{T} = (1,\,1,\,1,\dots)
\end{align}
This is a linear operation so the tangent linear action is identical to the forward model, and the derivative can immediately be written in terms of \(\mathbf{e}\).
\begin{align}\label{eq:reduce-derivatives}
   & \nabla\mathcal{R}(\delta\mathbf{x}) : \mathbf{X}\to\hat{X}
      = \mathbf{e}^{T}\delta\mathbf{x} \\
   & \nabla\mathcal{R}^{\star}(\delta\hat{x}^{\star}) : \hat{X}^{\star}\to\mathbf{X}^{\star}
      = \mathbf{e}\delta\hat{x}^{\star}
\end{align}
The adjoint of the reduction copies the adjoint input into all elements of the output, i.e. \(\nabla\mathcal{R}^{\star}=\mathcal{B}\) is a broadcast acting on \(\hat{X}^{\star}\).
Equivalently we can express the derivatives of broadcast in terms of a reduction acting on \(\mathbf{X}^{\star}\)
\begin{equation}\label{eq:bcast-derivatives}
   \mathcal{B}(\hat{x}) = \mathbf{e}\hat{x},
   \quad
   \nabla\mathcal{B} = \nabla\mathcal{R}^{\star},
   \quad
   \nabla\mathcal{B}^{\star} = \nabla\mathcal{R}.
\end{equation}
In order to maintain collective semantics, in practice the reduction is implemented using MPI's Allreduce so that all spatial comms hold a copy of the output.

\textbf{Shift}.
The last mapping is the shift operator needed for the time-halos in the model propagation \(\mathcal{L}\).
This mapping can be written as the matrix-vector product
\begin{equation}
   \mathcal{S}(\mathbf{x}) =
   \mathbf{E}\mathbf{x},
   \quad
   \mathbf{E} =
   \begin{pmatrix}
      0 &   &   &   \\
      1 & 0 &   &   \\
        & 1 & 0 &   \\
        &   & \ddots & \ddots
   \end{pmatrix}.
\end{equation}
Again this is linear so the tangent linear action and the adjoint can be written as
\begin{align}\label{eq:shift-derivatives}
   & \nabla\mathcal{S}(\delta\mathbf{x}) : \mathbf{X}\to\mathbf{X}
      = \mathbf{E}\delta\mathbf{x} \\
   & \nabla\mathcal{S}^{\star}(\delta\mathbf{x}^{\star}) : \mathbf{X}^{\star}\to\mathbf{X}^{\star}
      = \mathbf{E}^{T}\delta\mathbf{x}^{\star}
\end{align}
The action of \(\mathbf{E}^{T}\) is just a halo communication backward in time, consistent with the interpretation of the adjoint of a time-dependent model acting backwards in time.

The shift operation \(\mathcal{S}\) is used for the halo exchange in the model operator \(\mathcal{L}=(\mathcal{I}-\mathcal{T}_{\mathcal{M}}\circ\mathcal{S})\).
This halo exchange happens before the propagators \(\mathcal{M}\) are evaluated, e.g. if \(x_{i-1}\) is on spatial comm \(r-1\) and \(x_{i}\) is on spatial comm \(r\) then to calculate the misfit \(x_{i}-\mathcal{M}_{i}(x_{i-1})\) we first send \(x_{i-1}\) to spatial comm \(r\), which holds and evaluates \(\mathcal{M}_{i}\).
Looking at the structure of \(\mathbf{L}\) in Eq.~(\ref{methods:eq:L-block}) we can see that this is equivalent to each spatial comm ``owning'' the rows of \(\mathbf{L}\) corresponding to its local elements \(x_{i}\), which is analogous to the compressed-sparse-row (CSR) sparse matrix format.
The opposite is true of the adjoint model \((\mathcal{I} - \nabla\mathcal{S}^{\star}\circ\nabla\mathcal{T}_{\mathcal{M}}^{\star})\), where \(M_{i}^{T}x^{\star}_{i}\) is be calculated by spatial comm \(r\) before being sent backwards to spatial comm \(r-1\), which is analogous to the compressed-sparse-column (CSC) sparse matrix format.

The all-at-once operator \(\mathcal{L}\) is implemented with \texttt{AllAtOnceReducedFunctional}, which can be used to form \(\mathbf{L}\) and \(\mathbf{L}^{T}\).
On each spatial comm, a \texttt{ReducedFunctional} for each local \(\mathcal{M}_{i}\) is created during the \texttt{record\_stages} context manager in Listing \ref{lst:step-2} by taping the operations mapping the \texttt{stage.control} to the \texttt{state} passed to \texttt{set\_observation}.

The \texttt{WC4DVarReducedFunctional} makes the reduced functionals for each operation in Table \ref{tab:wc4dvar-mappings} available, which can be used for example to calculate the contributions of the model misfit \(\|\mathcal{L}(\mathbf{x})\|_{\mathbf{D}^{-1}}\) and observation misfit \(\|\mathcal{H}(\mathbf{x})\|_{\mathbf{R}^{-1}}\) to the total \(\mathcal{J}_{w}\) in Eq.~(\ref{methods:eq:wc-jhat}).

\subsubsection{PETSc/TAO coupling}\label{sec:software:impl:petsc}

\begin{table*}[t]
\caption{
   New matrix (\texttt{Mat}) and preconditioner (\texttt{PC}) types in Firedrake and Pyadjoint.
   The blocks of the \texttt{EnsembleBlockDiagonalMat} and \texttt{EnsembleBJacobiPC} can be any PETSc Mat type of the correct size.
   The \texttt{WC4DVarSchurPC}, \texttt{WC4DVarSaddlePC} and \texttt{AllAtOnceRFGaussSeidelPC} require a \texttt{ReducedFunctionalMat} constructed from the WC4DVar or all-at-once reduced functionals respectively.
   The \texttt{ReducedFunctionalMat} applies the action of the adjoint, tangent linear, or Hessian model, whereas previously only the Hessian action was available.
}
\label{tab:mats-pcs}
\renewcommand{\arraystretch}{1.25}
\begin{tabular}{cl}
\tophline
Mathematical object  & Software implementation \\
\middlehline
\(\mathbf{D}\), \(\mathbf{D}^{-1}\), \(\mathbf{R}\), \(\mathbf{H}\), \(\mathbf{H}^{T}\) & \texttt{EnsembleBlockDiagonalMat}, \texttt{EnsembleBJacobiPC} \\ \hline
\(\mathbf{\tilde{S}}^{-1}\) & \texttt{WC4DVarSchurPC} \\ \hline
\(\mathbf{A}\)              & \texttt{WC4DVarSaddlePC} \\ \hline
\(\mathbf{L}^{-1}\) & \texttt{AllAtOnceRFGaussSeidelPC} \\ \hline
\(\mathbf{\tilde{L}}^{-1}(M=I)\) & \texttt{AllAtOnceRFGaussSeidelPC} \\ \hline
\(\mathbf{L}\), \(\mathbf{L}^{T}\), \(M_{i}\), \(M_{i}^{T}\), \(H_{i}\), \(H_{i}^{T}\) & \texttt{ReducedFunctionalMat}  \\
\bottomhline
\end{tabular}
\end{table*}

In this section we describe how we achieve Aim 2 -- to automate the construction of the linear systems described in Sect. \ref{sec:methods:lsq} from \(\mathcal{J}_{w}\) -- through a set of PETSc matrix (\texttt{Mat}) and preconditioner (\texttt{PC}) types.
PETSc provides a variety of matrix and preconditioner types, but also allows the user to define custom types.
In Python this is done by creating a Python class (referred to as the Python context) that implements certain methods for PETSc to use as callbacks, for example a \texttt{Mat} context must have a \texttt{mult} method that implements matrix-vector products.
Firedrake makes heavy use of Python types for propagating symbolic information through the variational solvers \citep{kirby_solver_2018}.
The existing Pyadjoint Python Mat for applying the Hessian action of a \texttt{ReducedFunctional} has been extended to also provide Python contexts for the tangent linear and adjoint actions, which is essential for building the blocks of the saddle point system \(\mathbf{A}\) and the Schur complement approximation \(\mathbf{\tilde{S}}\).
The Mat and PC types used for the WC4DVar solver are shown in Table \ref{tab:mats-pcs} (excluding those for the covariance operators described in Sec \ref{sec:covariance}).

\textbf{Block diagonal matrices.}
The matrices \(\mathbf{D}\), \(\mathbf{D}^{-1}\), \(\mathbf{R}\), and \(\mathbf{H}\), and \(\mathbf{H}^{T}\) are all block diagonal with each block acting on a single component of an ensemble function space.
These are implemented with the \texttt{EnsembleBlockDiagonalMat} context which, on each spatial comm, requires a Mat for each local block.
These Mats can be of any type, for example in \(\mathbf{D}\) they are a \texttt{CovarianceMat}, and in \(\mathbf{H}\) they are a \texttt{ReducedFunctionalMat} for the tangent linear model of each \(\mathcal{H}_{i}\).
If the row and column spaces of \texttt{EnsembleBlockDiagonalMat} are the same size then it can be preconditioned with the \texttt{EnsembleBJacobiPC}, as shown in Listing \ref{lst:step-3-primal}.
On each spatial comm this creates a sub KSP for each local block, which can themselves be configured with any parameters compatible with the corresponding Mat (e.g. specifying the \texttt{CovariancePC} for blocks with a \texttt{CovarianceMat}).
When the \texttt{EnsembleBJacobiPC} is applied the blocks on different spatial comms are solved in parallel.

\textbf{Schur complement.}
For the approximate Schur complement \(\mathbf{\tilde{S}}\) the \texttt{WC4DVarSchurPC} sets up three KSPs, one for each of \(\mathbf{\tilde{L}}^{T}\), \(\mathbf{\tilde{D}}^{-1}\), and \(\mathbf{\tilde{L}}\).
The KSPs for the all-at-once tangent linear and adjoint models are each created from \(\mathbf{\mathcal{L}}\) using a \texttt{ReducedFunctionalMat}.
Rather than using an iterative method to approximately solve \(\mathbf{L}\) and \(\mathbf{L}^{T}\) it is common to simply replace them with one of the preconditioners \(\mathbf{\tilde{L}}\) described below.
For this reason the KSP type defaults to \texttt{preonly} and the user only has to specify a preconditioner (e.g. in lines 12-13 of Listing \ref{lst:step-3-primal}).
However, it is possible to instead use an iterative solver, although in this case care must be taken to ensure that \(\mathbf{\tilde{S}}\) is still symmetric.

\textbf{Saddle point system.}
For the \(3\times3\) saddle point system \(\mathbf{A}\) we can create the blocks using \texttt{ReducedFunctionalMat} or \texttt{EnsembleBlockDiagonalMat} as appropriate.
We use PETSc's MatNest type to combine the Mat for each block into a single Mat for \(\mathbf{A}\), which we can then create a KSP for in \texttt{WC4DVarSaddlePC}.
Because a MatNest inherently knows its block structure, it immediately works with the fieldsplit PC to give access to the Schur complement preconditioners \(\mathbf{P}_{U}\) etc.
Additionally, MatNest and the fieldsplit PC compose: when we tell PETSc to treat the \(3\times3\) \(\mathbf{A}\) as the \(2\times2\) \(\mathbf{\hat{A}}\) in Eq.~(\ref{eq:2x2-saddle-point}) it creates new MatNests for each of \(\mathbf{\hat{A}}_{00}\), \(\mathbf{\hat{A}}_{01}\), and \(\mathbf{\hat{A}}_{10}\).
We can then use the additive fieldsplit on \(\mathbf{\hat{A}}_{00}\) to solve \(\mathbf{D}\) and \(\mathbf{R}\) independently.
We could have created the \(2\times2\) MatNest for \(\mathbf{\hat{A}}\) ourselves so that we would not have to nest the fieldsplits, but providing the full \(3\times3\) structure gives the possibility to explore other preconditioning strategies in the future, for example forming the Schur complement on a different field.

\textbf{All-at-once system.}
Lastly, we describe the preconditioners for \(\mathbf{L}\) currently implemented.
The trivial ``preconditioner'' \(\mathbf{L}\left(M=0\right)~=~\mathbf{I}\) can be specified just using PETSc options by setting the PC type to \texttt{none}.
At the other extreme, \texttt{AllAtOnceRFGaussSeidelPC} applies the full inverse of \(\mathbf{L}\) (or \(\mathbf{L}^{T}\)) by stepping forwards (backwards) through the entire timeseries using the \(M_{i}\) (\(M_{i}^{T}\)) on each spatial comm in turn.
Whilst this gives no time parallelism it can be useful in two cases: as a comparison against approximations of \(\mathbf{L}^{-1}\); to emulate the forcing formulation of WC4DVar.
The approximation \(\mathbf{\tilde{L}}^{-1}\left(M=I\right)\) is structurally identical to \(\mathbf{L}^{-1}\), so we make this preconditioner available through \texttt{AllAtOnceRFGaussSeidelPC} by setting the \texttt{pc\_aaogs\_type} option to \texttt{identity}.
Although this also cannot be applied parallel-in-time, the computation at each stage is trivial.
We also consider approximations using fixed-bandwidth Richardson approximations to $\mathbf{L}$. We denote these by $\tilde{\mathbf{L}}^{-1}(R,2)$ for 2 iterations of Richardson, and $\tilde{\mathbf{L}}^{-1}(R,3)$ for three iterations of Richardson respectively.

\section{Numerical examples}\label{sec:examples}
In this section, we present numerical examples, with the goal of demonstrating that we have achieved the three aims of this work set out in Sect. \ref{sec:aims}.
Our first aim, A1, was to automate the construction of the cost function for a free choice of underlying differential equation and data assimilation system.
We present results for both SC4DVAR and WC4DVAR for Burgers' equation in Sect. \ref{sec:examples:Burgers results} and for WC4DVAR applied to the nonlinear shallow water equations in Sect. \ref{sec:examples:NLSWE}.
For the shallow water equations we consider different observation networks in Sect. \ref{sec:examples:changing observations}, illustrating the ease of changing the underlying data assimilation framework.
Our second aim, A2, was to  provide users with full flexibility to specify linear solver components.
We present results for a variety solver and preconditioner configurations for both examples, including the primal and saddle point formulations of the linear least-squares problem as presented in Sect. \ref{sec:methods:lsq}.
We also consider a number of approximations to $\mathbf{L}^T$ within the preconditioners in Sect. \ref{sec:examples:changing solver}.
Our final aim, A3, was to provide automatic space-time parallelism.
We demonstrate this scalability for the nonlinear shallow water equations in Sect. \ref{sec:examples:NLSWE}.
In each section we will illustrate which components of the Listings in Sect. \ref{sec:software} are changed to produce each set of experiments.
All other code remains unchanged. 

\subsection{Twin experiments and performance metrics}
In the following sections we  present a number of metrics that can be used to assess the performance of a data assimilation system.
We note that this paper does not seek to determine the best approach for a given problem, but instead to illustrate the tools available to users within the fdvar library.
We compute the value of the nonlinear objective function after the full assimilation cycle, as well as the gradient of the nonlinear objective function.
These metrics can be implemented for any data assimilation experiment, as they require no knowledge of a `truth'.
As discussed in Sect. \ref{sec:software:impl:autodiff}, we can also evaluate the model misfit $\|\mathcal{L}(\mathbf{x})\|_{\mathbf{D}^{-1}}$ and observation misfit $\|\mathcal{H}(\mathbf{x})\|_{\mathbf{R}^{-1}}$.

To validate our performance we implement a twin experiment, where we construct a `true' trajectory, and perturb this in order to obtain observations and a background state, as described in Sect. \ref{sec:covariance:implementation} and Alg. \ref{cov:alg:synthetic-data}.
We can then assess the fit of our analysis state, obtained after data assimilation, to the truth, and show how much  the assimilation has reduced the  error compared to the initial guess.
We compute  the norm of the error in the analysis compared to the truth at observation times, for physical variables, non-dimensional variables or unbalanced components.
We present results for the initial and final observation time, but the same approach can be applied at any point in the time-window.

The experiments below will be run for a fixed normalised wallclock time in order to compare the performance of different solver and preconditioner formulations for a prescribed computational budget.
As the computational cost of the optimisation routine is dominated by model evaluations, we define the normalised wallclock time by the number of model operations evaluated on a single processor.
We will indicate by $nproc$ the number of available parallel processors over which these model operations are distributed.

\subsection{Burgers' equation}\label{sec:examples:Burgers results}
We present results for the Burgers' problem introduced in Sect. \ref{sec:software:demo} for both SC4DVAR and WC4DVAR for $N_t=8$ observation times with observations being made every second model timestep. 
For WC4DVAR, we consider both the primal formulation as presented in Eq.(\ref{methods:eq:symmetricL1wc}) using Listing \ref{lst:step-3-primal}, and the saddle point formulation using Listing \ref{lst:step-3-saddle}.
In the latter case we present results for three choices of preconditioner, $\mathbf{P}_D,\mathbf{P}_L,\mathbf{P}_U$. This corresponds to changing line 16 in Listing \ref{lst:step-3-saddle} to \texttt{diagonal}, \texttt{lower} or \texttt{upper} respectively. 
Throughout this section we use the approximation $\tilde{\mathbf{L}}=\mathbf{L}$. For consistency, we present results for each problem according to the same normalised wallclock time across the optimisation.
We report results for two parallelisation frameworks: $nproc=1$ which represents a fully serial experiment, and $nproc=8$ which allows the $N_t=8$ observation times to be parallelised over $8$ processors.

\begin{table*}[t]
\caption{
Burgers' equation with SC4DVAR: Change in nonlinear objective function, Eq. (\ref{methods:eq:primalstrong}),  gradient, and error at the first and final observation times for initial guess and analysis. The linear solves are  preconditioned using Eq. (\ref{methods:eq:splitprecond}) with $\tilde{\mathbf{L}}=\mathbf{L}$. 
}
\label{examples:tab:BurgersSC}
\renewcommand{\arraystretch}{1.25}
\begin{tabular}{cllll}
\tophline
& $\mathcal{J}_s(x_0)$ & $\nabla  \mathcal{J}_s(x_0)$ & $\|x^{true}_0 - x_0\|$ & $\|x^{true}_8 - x_8\|$ \\
\middlehline
At initialisation & 12454 & 8140 &  $2.159\times 10^{-1}$& $3.523\times 10^{-1}$ \\ 
Analysis &  224.236 & 3.78882 & $1.458\times 10^{-2}$ & $2.366 \times 10^{-2}$\\
\bottomhline
\end{tabular}
\end{table*}

\begin{table*}[t]
\caption{Burgers' equation with WC4DVAR: Change in nonlinear objective function, Eq. (\ref{methods:eq:symmetricL1wc}),  gradient, and error at the first and final observation times for initial guess and analysis. We report results for the primal formulation, Eq. (\ref{methods:eq:symmetricL1wc}), preconditioned as in Eq. (\ref{methods:eq:symmetricL1wc}), and the saddle point formulation, Eq. (\ref{methods:eq:saddlepoint}) preconditioned with $\mathbf{P}_D$, Eq. (\ref{methods:eq:PD}), $\mathbf{P}_L$, Eq. (\ref{methods:eq:PL}) and $\mathbf{P}_U$, Eq. (\ref{methods:eq:PU}). In all cases we use $\tilde{\mathbf{L}}=\mathbf{L}$. }
\label{examples:tab:BurgersWC}
\renewcommand{\arraystretch}{1.25}
\begin{tabular}{clllll}
\tophline
Formulation&$nproc$ &  $\mathcal{J}_w(\mathbf{x})$ & $\nabla  \mathcal{J}_w(\mathbf{x})$ & $\|x^{true}_0 - x_0\|$ & $\|x^{true}_8 - x_8\|$ \\
\middlehline
At initialisation &- & 12454 & 632.453 &  $2.159\times 10^{-1}$& $3.523\times 10^{-1}$ \\ \hline
\multirow{2}*{Primal} & 1 & 205.265 &0.292763 &$1.186\times 10^{-2}$  & $1.656\times 10^{-2}$ \\
& 8 & 205.264 & 0.227249 & $1.187\times 10^{-2}$ & $1.657\times 10^{-2} $\\
\hline
\multirow{2}*{Saddle, $\mathbf{P}_D$} & 1 & 791.355 &46.5032  & $1.186\times 10^{-2}  $ & $2.266\times 10^{-2}$\\
& 8 & 205.265 & 0.246093  & $ 1.185\times 10^{-2}$ & $1.65\times10^{-2}$\\
\hline
\multirow{2}*{Saddle, $\mathbf{P}_L$} & 1 & 655.046 & 59.664 & $ 1.670\times 10^{-2} $ &$2.584\times 10^{-2}$\\
& 8 &  205.264 & 0.22724 & $1.187\times 10^{-2}$ &  $1.657\times 10^{-2}$\\
\hline
\multirow{2}*{Saddle, $\mathbf{P}_U$} & 1 &205.883 & 2.01284 & $1.17\times 10^{-2}$ & $4.743\times 10^{-2} $\\
& 8&205.264 & 0.22724 & $1.187\times 10^{-2}$ &  $1.657\times 10^{-2}$\\
\bottomhline
\end{tabular} 
\end{table*}

Table \ref{examples:tab:BurgersSC} shows results for SC4DVAR, and Table \ref{examples:tab:BurgersWC} reports results for the WC4DVAR, with a variety of formulations and preconditioners.
We report the value of the nonlinear objective function, the gradient and the error compared to the true trajectory at the first and final observation times. 
We see by exploiting the parallelisable structure of the saddle point formulation, we can obtain a lower objective function value for the same number of sequential model operators by distributing over 8 processors, and a lower gradient than solving the primal form. 
We note that the same truth trajectory and background are used for both SC4DVAR and WC4DVAR.
The flexibility of WC4DVAR allows us to achieve lower errors at both the initial and final observation time compared to solving SC4DVAR.
These results show the ease with which users can switch between data assimilation and solver options, and the variety of performance metrics that can be readily included within the fdvar library. 

\subsection{Nonlinear shallow water equations}\label{sec:examples:NLSWE}

In this section we present some experiments with the nonlinear rotating shallow water equations on the sphere, which we will use to demonstrate the use of: a variable transform as described in Sec. \ref{sec:covariance:mixed}; time-varying and nonlinear observation operators; different approximations for the all-at-once system \(\mathbf{L}\).
The shallow water equations are a canonical model of 2D atmospheric motion which are derived by assuming scale separation between vertical and horizontal motions in a thin atmosphere and vertically averaging the Euler equations to give a 2D velocity \(u\) tangent to the sphere and a depth \(h\).
\begin{equation}\label{examples:eq:swe}
\begin{aligned}
   \partial_{t}u + u\cdot\nabla u + fu^{\perp} & = -g\nabla h \\
   \partial_{t}h + \nabla\cdot\left(uh\right)  & = 0
\end{aligned}
\end{equation}
\(f\) is the Coriolis parameter, \(g\) is gravitaional acceleration, and \(u^{\perp}=e_{z}\times u\) with \(u\) the radial unit vector is a rotation by 90 degrees, defined on a the surface of a sphere of radius \(r\).
The shallow water equations are often used in the development of new atmospheric numerical models because they display several features of large scale atmospheric flow.
Of particular relevance is the splitting of slow balanced Rossby modes and fast unbalanced gravity wave modes.
Atmospheric flow is generally close to balance, so rather than using the state variable \(x=(u, h)\) as the control we will transform to \(z=(\varphi, \psi, h_{u})\) where the balanced and unbalanced parts of the full state are:
\begin{equation}\label{examples:eq:swe-decomposition}
\begin{aligned}
   u & = u_{u} + u_{b} = \nabla\varphi + \nabla^{\perp}\psi \\
   h & = h_{u} + h_{b} = h_{u} + \mathbb{H}\psi
\end{aligned}
\end{equation}
with a velocity potential \(\varphi\) defining the irrotational flow, a streamfunction \(\psi\) defining the solenoidal flow.
The balanced part of the flow is determined entirely by \(\psi\) with \(h_{b}=\mathbb{H}\psi\), where \(\mathbb{H}\) is defined using the linear balance relation \(fu_{b}^{\perp}=-g\nabla h_{b}\) as the solution of
\begin{equation}\label{examples:eq:linear-balance}
   \nabla\cdot\left(f\nabla\psi\right) = -g\nabla^{2}h_{b}
\end{equation}
Defining \(D\) and \(G\) as the discretisations of \(\nabla\) and \(\nabla^{\perp}\) respectively, the decomposition Eq.~(\ref{examples:eq:swe-decomposition}) can be written as a control variable transform \(x=Kz\):
\begin{equation}\label{examples:eq:cvt}
   \begin{pmatrix}
      u \\ h
   \end{pmatrix}
   =
   \begin{pmatrix}
      D & G          &   \\
        & \mathbb{H} & I \\
   \end{pmatrix}
   \begin{pmatrix}
      \varphi \\ \psi \\ h_{u}
   \end{pmatrix}
\end{equation}
Assuming that the components of \(z\) are uncorrelated with each other, then the full covariance operator implied by Eq.~(\ref{examples:eq:cvt}) is:
\begin{equation}\label{bal:eq:full-covariance}
   B = KB_{z}K^{T} =
   \begin{pmatrix}
      DB_{\varphi}D^{T} + GB_{\psi}G^{T} & GB_{\psi}\mathbb{H}^{T} \\
      \mathbb{H}B_{\psi}G^{T}            & B_{h} + \mathbb{H}B_{\psi}\mathbb{H}^{T}
   \end{pmatrix}
\end{equation}
This is just one possible analytical balance relation, see \citet{bannisterBalanceConditionsVariational2021} for other possible balance operators.
The shallow water equations Eq.~(\ref{examples:eq:swe}) and the transform Eq.~(\ref{examples:eq:cvt}) are discretised in space using compatible finite elements \citep{gibson_compatible_2019} and in time using the implicit midpoint rule using the Irksome library for implicit Runge-Kutta methods \citep{farrell_irksome_2021} using Firedrake.

The experimental setup is based on the second test case from \citet{williamson_standard_1992} of solid body rotation in geostrophic balance, with a maximum velocity of \(40m/s\).
The ground truth trajectory for each example is then generated using the method described above.
A larger lengthscale and variance is used for \(\psi\) in \(B\) to maintain approximate balance in the flow.

We first present results for a base case with an assimilation window of 4 hours with $5$ observation times evenly spaced every 0.8 hours, with point observations taken at every second model timestep.
As in Section \ref{sec:examples:Burgers results}, we fix the budget for each linear system solve.
We first present results in Table \ref{examples:tab:SWEWC} results for WC4DVAR with an assimilation window of 4 hour and considers $N_t=5$, with point observations taken every 0.8 hours at every second model timestep.
We take point observations of both the velocity and depth using a grid of 32 latitude and 32 longitude locations with a maximum latitudinal location of $\pm 0.4\pi$ radians.
As in Sect. \ref{sec:examples:Burgers results}, we fix the budget for each linear system solve.
For consistency, we present results for each experiment according to the same normalised wallclock time.
We will build upon this `base' case in the remainder of this section. 
Table \ref{examples:tab:SWEWC} show results for different formulations and preconditioners applied to  WC4DVAR  using $\tilde{\mathbf{L}} = \mathbf{L}$.
We report the nonlinear objective function and its gradient, in addition to the objective misfit and its gradient, and the model misfit and its gradient.
We see that the value of  $\mathcal{J}_m(\mathbf{x})$ is very small at initialisation, which results from using a smooth background term.
For all  formulations and preconditioners the analysis reduces the value of the objective function and improves the observation misfit compared to the initial condition.

\begin{table*}[t]
\caption{
Nonlinear shallow water equation with WC4DVAR: Change in nonlinear objective function, Eq. (\ref{methods:eq:symmetricL1wc}),  gradient and corresponding observation and model misfits and gradients. We report results for the primal formulation, Eq. (\ref{methods:eq:symmetricL1wc}), preconditioned as in Eq. (\ref{methods:eq:symmetricL1wc}), and the saddle point formulation, Eq. (\ref{methods:eq:saddlepoint}) preconditioned with $\mathbf{P}_D$, Eq. (\ref{methods:eq:PD}), and $\mathbf{P}_L$, Eq. (\ref{methods:eq:PL}). In all cases we use $\tilde{\mathbf{L}}=\mathbf{L}$.}
\label{examples:tab:SWEWC}
\renewcommand{\arraystretch}{1.25}
\begin{tabular}{cllllll}
\tophline
Formulation& $\mathcal{J}_w(\mathbf{x})$ & $\nabla  \mathcal{J}_w(\mathbf{x})$ &$\mathcal{J}_o(\mathbf{x})$ & $\nabla  \mathcal{J}_o(\mathbf{x})$ &$\mathcal{J}_m(\mathbf{x})$ & $\nabla  \mathcal{J}_m(\mathbf{x})$  \\
\middlehline
At initialisation & $1.215\times10^5$ &$6.457\times 10^3$ &  $1.215\times10^5$  &$6.457\times10^3$&$2.210\times 10^{-10}$&$1.252\times10^{-2}$\\ 
{Primal} & $4.451\times10^4$ & $1.564\times10^5$&$9.145\times10^3$ & $3.199\times10^2$&$3.536\times10^4$& $1.564\times10^5$  \\
{Saddle, $\mathbf{P}_D$} & $3.274\times10^4$ & $9.644\times10^4$ & $1.115\times 10^4 $ & $3.375\times 10^2$ & $2.159\times 10^4$& $9.644\times 10^4$\\
{Saddle, $\mathbf{P}_L$} &$3.588\times10^4$ & $1.203\times10^5$ & $9.267\times 10^3 $ & $2.498\times10^2$ & $2.662\times10^4$ & $1.204\times 10^5$\\
\bottomhline
\end{tabular} 
\end{table*} 

\subsubsection{Changing the objective \(J\) (observation variations)}\label{sec:examples:changing observations}
We now present results which alter the observation network by changing the definition of the observation operator (similar to lines 25-32 of Listing \ref{lst:step-2}).
Table \ref{examples:tab:SWEWC_changeobfreqtime} presents results where the observation frequency and observation density have been altered compared to the observation network presented in Table \ref{examples:tab:SWEWC}.
We consider two settings where the total number of observations remains fixed over the observation window.
Firstly, we reduce the observation density in latitude only to $16$, and increase the observation frequency to every model timestep.
The observation time window is the same as in the base case.
The second observation network maintains the same point density as the base case but incorporates observation at twice the information every model timestep.
This results in an observation window that is half as long as for the results in Table \ref{examples:tab:SWEWC}.
In all cases $N_t=5$.

\begin{table*}[t]
\caption{
Nonlinear shallow water equation under changing observation network: Change observation misfit, observation misfit  gradient, and error at the final observation times after data assimilation is applied. We report results for the primal formulation, Eq. (\ref{methods:eq:symmetricL1wc}), preconditioned as in Eq. (\ref{methods:eq:symmetricL1wc}), and the saddle point formulation, Eq. (\ref{methods:eq:saddlepoint}) preconditioned with $\mathbf{P}_D$, Eq. (\ref{methods:eq:PD}) and  $\mathbf{P}_L$, Eq. (\ref{methods:eq:PL}) all using $\tilde{\mathbf{L}}=\mathbf{L}.$}
\label{examples:tab:SWEWC_changeobfreqtime}
\renewcommand{\arraystretch}{1.25}
\begin{tabular}{cllllll}
\tophline
 &Formulation&  $\mathcal{J}_o(\mathbf{x})$ &$\nabla \mathcal{J}_o(\mathbf{x})$& $\|\phi^{true}_{end} - \phi_{end}\|$ & $\|\psi^{true}_{end} - \psi_{end}\|$ & $\|(h_u)_{end}^{true}- (h_u)_{end}\|$  \\
\middlehline
  &At initialisation   &$1.127\times10^5$& $3.286\times10^3$&$4.60\times10^{-3}$&$3.59\times10^{-2}$ & $5.01\times10^{-3}$\\ 
nlat = 16,&{Primal} & $8.567\times10^3$  &$2.186\times10^2$ & $1.22\times10^{-3}$& $3.35\times10^{-3}$ & $2.01\times10^{-3}$\\
tend = 4&{Saddle, $\mathbf{P}_D$} &$1.008\times10^4$  &$2.472\times10^2$ & $1.51\times10^{-3}$ & $3.33\times10^{-3}$& $2.39\times10^{-3}$\\
&{Saddle, $\mathbf{P}_L$} & $8.390\times10^3$ & $2.007\times10^2$& $1.31\times10^{-3}$& $1.76\times10^{-3}$ & $2.15\times10^{-3}$\\
\hline
&At initialisation   &$1.226\times10^5$ & $6.432\times10^3$&$2.27\times10^{-3}$&$3.68\times10^{-2}$& $3.90\times10^{-3}$\\ 
nlat = 32,&{Primal} & $9.598\times10^3$  &$3.320\times10^2$ & $1.02\times10^{-3}$ & $2.10\times10^{-3}$& $2.03\times10^{-3}$\\
 tend = 2 &{Saddle, $\mathbf{P}_D$} & $1.133\times10^4$ &$3.056\times10^2$ &  $1.30\times10^{-3}$&$1.92\times10^{-3}$& $2.49\times10^{-3}$\\
&{Saddle, $\mathbf{P}_L$}& $9.646\times10^3$& $2.346\times10^2$ & $1.16\times10^{-3}$&$1.27\times10^{-3}$&$2.25\times10^{-3}$\\
\bottomhline
\end{tabular} 
\end{table*}

In Table \ref{examples:tab:SWEWC_changeobfreqtime} we report the observation misfit, $\mathcal{J}_o(\mathbf{x})$, and gradient of the observation misfit, $\nabla \mathcal{J}_o(\mathbf{x})$, as well as the fit to the control variables at the final observation time.
We note that the final observation time is $t=10$ for Case 1, and $t=5$ for Case 2.
All data assimilation approaches reduce both the observation misfit and its gradient by at least an order of magnitude, with gradient reductions being larger for the second observation framework.
This results in improved fit to the control variables; all problem formulations are able to reduce the error in $\phi_{end}$ by an order of magnitude compared to the initial guess.
In both cases solving the primal form leads to larger reductions in the errors for $\phi$ and $h_u$, but larger errors in $\psi$ than solving the saddle point formulation with either block diagonal or block triangular preconditioner.
It is also straightforward to consider errors at each observation time in terms of physical variables, allowing flexibility for users to assess their performance against the most relevant metrics for their setting. 

\begin{table*}[t]
\caption{
Nonlinear shallow water equation for nonlinear observation operator: Change in the observation misfit,  and error at the final observation times after data assimilation is applied. We report results for the primal formulation, Eq. (\ref{methods:eq:symmetricL1wc}), preconditioned as in Eq. (\ref{methods:eq:symmetricL1wc}), using $\tilde{\mathbf{L}}=\mathbf{L}.$
}
\label{examples:tab:SWEWC_changeobsoperator}
\renewcommand{\arraystretch}{1.25}
\begin{tabular}{clllll}
\tophline
Point observations &  $\mathcal{J}_o(\mathbf{x^{(0)}})$ & $\mathcal{J}_o(\mathbf{x^{(3)}})$ &  $\|\psi^{true}_{end} - \psi_{end}\|$ & $\|u^{true}_{end} - u_{end}\|$ & $\|h_{end}^{true}- h_{end}\|$  \\
\middlehline
{\(20\times20\)/\(0.4hr\)}  & \(2.692\times10^5\) & \(1.005\times10^4\) & \(3.06\times10^{-3}\) & \(3.00\times10^{-2}\) & \(2.70\times10^{-3}\) \\
\hline
{\(28\times 29\)/\(0.8hr\)} & \(1.196\times10^5\) & \(1.113\times10^4\) & \(2.47\times10^{-3}\) & \(2.58\times10^{-2}\) & \(2.65\times10^{-3}\) \\
\hline
{\(40\times40\)/\(1.6hr\)}  & \(1.324\times10^5\) & \(1.242\times10^4\) & \(2.43\times10^{-3}\) & \(2.61\times10^{-2}\) & \(2.72\times10^{-3}\) \\
\bottomhline
\end{tabular} 
\end{table*}

Table \ref{examples:tab:SWEWC_changeobsoperator} reports results when incorporating a second, nonlinear choice of observation operator, i.e. including additional definitions after line 33 of Listing \ref{lst:step-2}.
Here we consider observations taken on latitudinal lines \(\Gamma_{j}\) with endpoints at latitudes \(\pm 0.25\pi\)
The depth and zonal velocity are interpolated onto these lines, and two types of observations are taken.
Firstly, the 1D profile of the depth along each line \(\Gamma_{j}\).
Secondly, the total zonal mass flux \(\dot{m}_{j}\) through each line, defined as the integral of the mass flux \(u_{z}h\) over the line.
\begin{equation}\label{examples:eq:mass-flux}
   \dot{m}_{j} = \int_{\Gamma_{j}} u_{z}h\mathrm{d}s
\end{equation}
The tangent linear and adjoint models are automatically derived for the interpolation, product, and integration, just as for other operations.
The assimilation window is extended to 4.8 hours, keeping the timestep size at 0.4 hours.
In all three cases in Table \ref{examples:tab:SWEWC_changeobsoperator}, 10 lines are used, equally spaced longitudinally around the sphere, and line observations are taken every 1.6 hours. 
In each case the density and frequency of the point observations is varied, again keeping the total number of point observations over the assimilation window approximately fixed: \(20\times20\) points observed every 0.4 hours; \(28\times29\) points observed every 0.8 hours; \(40\times40\) points observed every 1.6 hours.
The variational problem is solved using the primal formulation with \(\tilde{\mathbf{L}}=\mathbf{L}\), and 4 inner iterations in each of the 3 outer iterations.

Because the observation network is different for each case the initial observation misfit \(\mathcal{J}_{o}(\mathbf{x}^{(0)})\) also varies.
In all three cases the observation misfit has been decreased by at least an order of magnitude after 3 outer iterations (\(\mathcal{J}_{o}(\mathbf{x}^{(3)})\)).
We also show the errors in the streamfunction, total velocity, and total height at the end of the assimilation window in each case.
The total height errors are similar between each of the three cases, however the streamfunction and velocity errors are around a fifth higher in the case with point observations every 0.4 hours.
This is potentially because, although there are approximately the same total number of observations as the other two cases, the points distributed more sparsely, and the line observations provide more information about the height field that the velocity field.

The results in this section demonstrate the ease with which it is possible to alter the observation network, within the fdvar library, including integrals, nonlinear observation operators, and combining observing networks at each observation time, thus achieving Aim 1. 

\subsubsection{Changing the solver (iterative method or \(M\) approximations)}\label{sec:examples:changing solver}
In this section we present results that illustrate the flexibility in solver options that can be applied to the linear system, i.e. Aim 2.
Our results focus on changes to  $\tilde{\mathbf{L}}$ within the preconditioners that have been considered in previous sections.
We achieve this by altering lines 13-14 in Listing \ref{lst:step-3-primal} for the primal form and line 26 in Listing \ref{lst:step-3-saddle} for the saddle point formulation. 
Further flexibility  is immediately available courtesy of the TAO library.
It is straightforward to alter the nonlinear and linear solvers, for-example replacing standard Newton with a line-search on line 4 of Listings \ref{lst:step-3-primal} and \ref{lst:step-3-saddle}, or  Gauss-Newton by changing line 3 of Listings \ref{lst:step-3-primal} and \ref{lst:step-3-saddle}, as is typically used in meteorological applications of 4DVar.
Instead of applying flexible CG to the linear solve in the primal system, alternative solvers such as GMRES can be used by changing line 6 in Listing \ref{lst:step-3-primal}.
Preconditioners for the linear system solve can be applied on both the left and right simply by changing an option in the TAO options file.

Table \ref{examples:tab:NLSWEDiffLhat} presents results for the nonlinear shallow water equations for three of choices of $\tilde{\mathbf{L}}$ introduced in Sect. \ref{sec:software:impl:petsc};  replacing model operators in $\mathbf{L}$ with the identity, denoted $\tilde{\mathbf{L}}^{-1}(M=I)$, approximating $\mathbf{L}$ with two iterations of a Richardson iteration, denoted $\tilde{\mathbf{L}}^{-1}(R,2)$,  and approximating $\mathbf{L}$ with three iterations of a Richardson iteration, denoted $\tilde{\mathbf{L}}^{-1}(R,3)$.
Each of these choices result in more parallelisable preconditioners compared to the choice of $\mathbf{L}$ used in the previous section, which can be exploited within the fdvar framework. For all formulations, we active the lowest objective function value when applying $\tilde{\mathbf{L}}^{-1}(R,3)$, with the smallest gradient occuring for $\tilde{\mathbf{L}}^{-1}(M=I)$. 

\begin{table*}[t]
\caption{
Nonlinear SWE equation with changing preconditioners: Value of nonlinear objective function, Eq. (\ref{methods:eq:symmetricL1wc}) and gradient after data assimilation is applied. We report results for the primal formulation, Eq. (\ref{methods:eq:symmetricL1wc}), preconditioned as in Eq. (\ref{methods:eq:symmetricL1wc}), and the saddle point formulation, Eq. (\ref{methods:eq:saddlepoint}) preconditioned with $\mathbf{P}_D$, Eq. (\ref{methods:eq:PD}), and  $\mathbf{P}_L$, Eq. (\ref{methods:eq:PL}) for a variety of approximations to $\tilde{\mathbf{L}}$.
}
\label{examples:tab:NLSWEDiffLhat}
\renewcommand{\arraystretch}{1.25}
\begin{tabular}{clll|lll}
\hline
Formulation&\multicolumn{3}{c|}{Nonlinear objective function } &  \multicolumn{3}{c}{Gradient} \\
 &   {$\tilde{\mathbf{L}}^{-1}(M=I)$} &  {$\tilde{\mathbf{L}}^{-1}(R,2)$} &  {$\tilde{\mathbf{L}}^{-1}(R,3)$}  &  {$\tilde{\mathbf{L}}^{-1}(M=I)$} &  {$\tilde{\mathbf{L}}^{-1}(R,2)$} &  {$\tilde{\mathbf{L}}^{-1}(R,3)$} \\
\hline

{Primal} &$1.179\times10^5$ &$1.107\times10^5$&$4.686\times 10^4$&$9.422\times 10^3$&$3.591\times 10^5$& $2.004\times10^5$\\

{Saddle, $\mathbf{P}_D$} & - &$4.254\times 10^4$&$3.843\times 10^4$&-&  $1.493\times10^5$&$1.279\times10^5$\\

{Saddle, $\mathbf{P}_L$} &  $1.188\times10^5$& $4.090\times10^4$&$3.901\times10^4$&$1.803\times10^3$&$1.673\times10^5$& $1.428\times 10^5$ \\

\hline
\end{tabular} 
\end{table*}

It is similarly straightforward to approximate other terms within the preconditioners, for example using $\tilde{\mathbf{D}}$ that uses fewer iterations of the diffusion operator by changing lines 16-21 in Listing \ref{lst:step-3-primal} or line 27 in Listing \ref{lst:step-3-saddle}.
This may also be desirable in the case of more complicated observation error covariance matrices.
Our results consider a fixed normalised computational budget, with a constant number of iterations of the nonlinear Newton method.
Allowing full control over the balance between nonlinear and linear iterations, and stopping criteria will allow users to more fully explore and test the full extent of optimisation methods more readily than in the past. 

\subsubsection{Scaling study}

Parallel scaling results are shown in tables \ref{examples:tab:Jhat-timing} and \ref{examples:tab:solve-timing}.
Wallclock times are calculated on a local workstation with an ADM Ryzen Threadripper 5975WX.
The timed assimilation window is 6 hours, with a timestep of 0.5 hours and 6 observation times, for a total of 12 model timesteps.
The window is run in serial, and in parallel with 2, 3, or 6 ranks in time, giving a total of 6, 3, 2, or 1 observation stage per spatial comm.
Table \ref{examples:tab:Jhat-timing} shows the wallclock time to re-evaluate the objective functional, as well as its adjoint and tangent linear models, after the recording is finished.
We can see that the timings scale very well as the time-parallelism is increased, as shown by the wallclock time per observation stage on each spatial communicator, which is a measure of the total "sequential" work.
Table \ref{examples:tab:Jhat-timing} shows the wallclock time to complete a single outer iteration with three inner iterations of the primal formulation, using 3 different preconditioners with varying degrees of parallelism.
The modelled sequential cost is shown in brackets next to the timing for each case.
This modelled cost for the perfect model \(\mathbf{L}\) shows the limited theoretical scalability of this preconditioner compared to the preconditioners with approximate \(\tilde{\mathbf{L}}\).
The times scale reasonably well with the expected cost, although some variation in the agreement is seen between the different cases.
For example the timing for the perfect model \(\mathbf{L}\) follows the modelled cost much more closely than the identity model \(\tilde{\mathbf{L}}(M=I)\) timing.
This variation is to be expected given that we have neglected all operations except the \(\mathcal{M}\) evaluations from the cost model.
In future work we will profile the performance on HPC facilities, and further develop cost modelling to include other parts of the algorithm.

\begin{table*}[t]
\caption{Wallclock time to evaluate the forward, tangent linear, and adjoint models on increasing number of processors. Wallclock time per local observation stage shown in brackets.}
\label{examples:tab:Jhat-timing}
\renewcommand{\arraystretch}{1.25}
\begin{tabular}{cccc}
\tophline
$nproc$ &  $\mathcal{J}$ & $\nabla\mathcal{J}$ & $\nabla\mathcal{J}^{\star}$ \\
\middlehline
1       & 7.0 (1.2)      & 7.7 (1.3)           & 7.9 (1.3)                   \\ \hline
2       & 3.9 (1.3)      & 4.3 (1.4)           & 4.4 (1.5)                   \\ \hline
3       & 2.8 (1.4)      & 3.2 (1.6)           & 3.2 (1.6)                   \\ \hline
6       & 1.8 (1.8)      & 2.1 (2.1)           & 1.9 (1.9)                   \\
\bottomhline
\end{tabular} 
\end{table*}

\begin{table*}[t]
\caption{Wallclock time to calculate one outer iteration and three inner iterations with the primal formulation using different preconditioners \(\tilde{\mathbf{L}}^{-1}\) for \(\mathbf{L}^{-1}\) on increasing number of processors. Modelled cost in number of sequential model \(\mathcal{M}\) integrations in brackets.}
\label{examples:tab:solve-timing}
\renewcommand{\arraystretch}{1.25}
\begin{tabular}{ccccc}
\tophline
$\tilde{\mathbf{L}}^{-1}$ &  $nproc=1$  & $nproc=2$  & $nproc=3$  & $nproc=6$ \\
\middlehline
\(\mathbf{L}^{-1}\)       & 195.1 (120) & 127.9 (84) & 104.2 (72) &  83.5 (60) \\ \hline
\(\mathbf{L}^{-1}(M=I)\)  & 152.3 (72)  &  85.5 (36) &  62.2 (24) &  40.5 (12) \\ \hline
\(\mathbf{L}^{-1}(R,2)\)  & 363.8 (168)  & 259.6 (84) & 221.8 (56) & 189.9 (28) \\
\bottomhline
\end{tabular} 
\end{table*}

\conclusions\label{sec:conclusions} 
We have presented fdvar, a Python library which allows users to apply variational data assimilation methods to a wide array of problems, with full flexibility, and minimal developer overhead. We achieve this by making use of Firedrake's automatic differentiation capabilities, which ensure that once users have implemented the forward problem, gradient and adjoint information is available immediately. This allows us to construct the resulting nonlinear parallel-in-time objective functional from a limited number of user inputs which define the underlying data assimilation problem. 
 In addition to the parallelisable framework offered by fdvar, we also provide implementations of autoregressive covariance operators, and basic observation operators that are compatible with the finite element framework of Firedrake.
 
 We illustrate these via a code demo applied to Burgers' equation, showing the succinct and legible manner in which users can specify all data assimilation and solver components. This walkthrough also allows us to illustrate the parallelism that is embedded in the construction of the 4DVar problem, which is automatically exploited in order to accelerate computational performance. We show how the syntax can readily be adapted to different problem formulations, and nested to permit increasing levels of complexity within preconditioners, without adding complexity to the parameter inputs. 

We illustrate the modularity and usability of the library via two numerical examples, presenting results for the Burgers' equation, and more comprehensive investigations for the nonlinear shallow water equations. We present results for a variety of observation networks, including a combination of point and integral observations, and show that different performance metrics are straightforward to compute, providing additional insight at minimal additional computational cost. We  consider different approximations to model operators within preconditioners. Finally, we present a scaling study which confirms space-time parallelism is available immediately courtesy of the fdvar library. 

 The fdvar library achieves the following three aims, namely to
\begin{enumerate}
   \item[A1.] Automate the construction of the nonlinear objective function while allowing the user full control over the specification of its components.
   \item[A2.] Automate the construction of the resulting linear systems within the optimisation routine, while allowing the user full control over the specification of the linear solver.
   \item[A3.] Provide performant space-time parallelism without requiring the user to coordinate parallel communications.
\end{enumerate}
With this, users are now able to test variational data assimilation methods on new problems, with minimal computational development overhead. This removes a major bottleneck in stress-testing and validating existing and new approaches, including in testing optimisation and numerical linear aspects relating to solvers and preconditioners. 
While variational data assimilation methods have seen huge success within geosciences, the need to develop and maintain adjoint codes has represented a significant barrier in their application to new fields.  The fdvar library eliminates this requirement, opening up variational data assimilation methods to a much wider range of scientific research fields.

\codedataavailability{
All code used in this manuscript is free and open source.
The fdvar library is available from
https://github.com/firedrakeproject/fdvar and is released under the MIT license.

}

\authorcontribution{Conceptualisation: JHC, DAH, JMT. Data curation: JHC. Formal analysis: JHC, CC. Funding acquisition: DAH, JMT. Investigation: JHC. Methodology: JHC, DAH, JMT. Project administration: DAH. Software: JHC, DAH, JMT.  Validation: JHC, JMT. Visualisation: JHC. Writing (original draft preparation): JHC, JMT. Writing (review and editing): JHC, DAH, JMT.} 

\competinginterests{DAH is the chief-executive editor of Geoscientific Model Development} 

\begin{acknowledgements}
The authors would like to thank Jack Betteridge and James Maddison respectively for the PyOP2 and PETSc implementations of the mass matrix square root method in section \ref{sec:covariance}.
JHC would like to thank Colin Cotter and James Maddison for many useful discussions during this work.

This work was supported by the Engineering and Physical Sciences Research Council (EP/W026066/1) and the Computational Science Centre for Research Communities (CoSeC) Fellowship programme. The authors wish to thank the attendees of the Data Assimilation in Firedrake Hackathon for the useful discussions and suggestions. 
\end{acknowledgements}

\bibliographystyle{copernicus}
\bibliography{fdvar.bib}

\end{document}